\documentclass[a4paper, 10pt]{article}
\usepackage[paper=a4paper, left=1.0in, right=0.8in, top=1.0in, bottom=1.0in]{geometry}
\usepackage{amsmath}  
\usepackage{alltt}    
\usepackage{graphicx} 
\usepackage{appendix}
\usepackage{subfig}
\usepackage{rotating}
\usepackage{setspace}
\usepackage{verbatim}
\usepackage{listings}
\usepackage{color}
\usepackage{footmisc}
\usepackage{url}
\usepackage{hyperref}
\usepackage{multirow}
\usepackage{caption}
\usepackage{float}
\usepackage{color}
\usepackage{pifont} 
\begin{document}

\title{\Large\bf MAIL: Malware Analysis Intermediate Language}
{
\small
\author
{
	\bf Shahid Alam \\
	\textit{Department of Computer Science} \\
	\textit{University of Victoria, BC, V8P5C2} \\
	\textit{E-mail: salam@cs.uvic.ca} \\
	\today
}
}
\date{}
\maketitle
\thispagestyle{empty}

\section{Introduction}\label{sec:introduction}

Intermediate languages are used in compilers \cite{Dragon-Book} to translate the source code into a form that is easy to optimize and increase portability. The term intermediate language also refers to the intermediate language used by the compilers of high level languages that does not produce any machine code, such as Java and C\#. An example of adding two numbers in the intermediate language CIL (Common Intermediate Language) used in implementing C\# is as follows:

{
\small
\begin{verbatim}
   a = a + b;
   is translated to the following CIL code:
   ldloc.0        ; Push the first local on the stack
   ldloc.1        ; Push the second local on the stack
   add            ; Pop the two locals, add them and push the result on the stack
   stloc.0        ; Pop the result and store it in the first local
\end{verbatim}
}

CIL is a stack based language, i.e: the data is pushed on the stack instead of pulling from the registers. That is one of the reasons why, in the example above, one simple add statement is translated into four stack-based statements. The same add statement can be translated into the \emph{three address code} \cite{Dragon-Book} as:

{
\small
\begin{verbatim}
   a := a + b
\end{verbatim}
}

The \emph{three address code} is an intermediate language used by most of the compilers. The two popular open source compilers GCC \cite{GCC} and LLVM \cite{LLVM} use \emph{three address code} as part of there intermediate languages.

\subsection{Hidden Malwares (Obfuscation)}

Detecting whether a given program is a malware (virus) is an undecidable problem \cite{Virus-Undecidable-Problem-1, Virus-Undecidable-Problem-2}. Antimalware software detection techniques are limited by this theoretical result. Malware writers exploit this limitation to avoid detection.

In the early days the malware writers were hobbyists but now the professionals have become part of this group because of the financial gains \cite{ITU-Malware-Financial} attached to it. One of the basic techniques used by a malware writer is obfuscation \cite{Obfuscation-2}. Such a technique obscure a code to make it difficult to understand, analyze and detect malwares embedded in the code.

Initial obfuscators were simple and were detected by signature-based detectors. These signature-based detectors work on simple signatures such as byte sequences, instruction sequences and string signatures (pattern of a malware that uniquely identifies it). They lack information about the semantics or behavior of the malicious program. To counter these detectors the obfuscation techniques evolved. Some of the mutations used in polymorphic and metamorphic \cite{Obfuscation-1} malwares are:

\begin{itemize}

   \item
\textbf{Instruction reordering:} By changing the ordering of instructions with commutative or associative operators, the structure of the instructions can be changed. This reordering does not change the behavior of the program. As a simple example:

{
\small
\begin{verbatim}
a = 10; b = 20;                   a = 10; b = 20;
x = a * b;   can be changed to:   x = b * a

original machine code    and       assembly:

c7 45 f4 0a 00 00 00 	   movl   [rbp-0xc], 0xa  ; a = 10
c7 45 f8 14 00 00 00 	   movl   [rbp-0x8], 0x14 ; b = 20
8b 45 f4             	   mov    eax, [rbp-0xc]  ;
0f af 45 f8          	   imul   eax, [rbp-0x8]  ; a * b
89 45 fc             	   mov    [rbp-0x4], eax  ; x = a * b

changed machine code     and       assembly:

c7 45 f4 0a 00 00 00 	   movl   [rbp-0xc], 0xa  ; a = 10
c7 45 f8 14 00 00 00 	   movl   [rbp-0x8], 0x14 ; b = 20
8b 45 f8             	   mov    eax, [rbp-0x8]  ; (reordered)
0f af 45 f4              imul   eax, [rbp-0xc]  ; b * a (reordered)
89 45 fc                 mov    [rbp-0x4], eax  ; x = b * a
\end{verbatim}
}

Because of the two reordered instructions the original and the changed machine codes have different signatures. Other instructions can also be reordered if no dependency exists between the instructions.
   \item
\textbf{Dead code insertion:} Dead code is a code that either does not execute or has no effect on the results of a program. Following is an example of dead code insertion:
{
\small
\begin{verbatim}
mov   ebx, [ebp+4]
add   ebx, 0x0           ; dead code
nop                      ; dead code
jmp   ebx
\end{verbatim}
}
   \item
\textbf{Register renaming:} To avoid detection registers are reassigned in a fragment of a binary code. This changes the byte sequence (signature) of the machine code. A signature-based detector will not be able to match the signature if it is searching for a specific register. An example of register renaming is given below (register \emph{eax} is renamed to \emph{edx}):
{
\small
\begin{verbatim}
lea   eax, [RIP+0x203768]               lea   edx, [RIP+0x203768]
add   eax, 0x10                         add   edx, 0x10
jmp   eax                               jmp   edx
\end{verbatim}
}
   \item
\textbf{Order of instructions:} To change the control flow of a program the order of instructions is changed in the binary image of the program, keeping the order of execution the same by using jump instructions.
   \item
\textbf{Branch functions:} A branch function is used \cite{Obfuscation-2} to obscure the flow of control in a program. The target of all or some of the unconditional branches in a program is replaced by the address of a branch function. The branch function makes sure the branch is correctly transferred to the right target for each branch.
\end{itemize}

Because of the financial gains attached with the malware industry, malware writers are always targeting new technologies. To improve the detection of malwares, especially metamorphic malwares, we need to develop new methods and techniques to analyze behavior of a program, to make a better detection decision with few false positives.

\subsection{Why an Intermediate Language for Malware Analysis}

Here we are going to list some of the reasons why we need to transform a program in an assembly language to an intermediate language for malware analysis:

\begin{itemize}
\item
There are hundreds of different instructions in any assembly language. For example the number of instructions in the two most popular ISAs (Instruction Set Architectures) are: Intel x86-64 = 800+ \cite{Intel-Developer-Manual} and ARM = 400+ \cite{ARM-Architecture-Manual}. We need to reduce the number of these instructions considerably to optimize the static analysis of any such assembly program.
\item
Not only the different instructions are big in numbers but they are also big in complexity, such as Intel x86-64 instruction's \emph{PREFETCHh}, \emph{MOVD} and \emph{MOVQ}. The instruction \emph{PREFETCHh} moves data from the memory to the cache. Is this action important, if we are performing a static analysis for detecting malwares? Our answer is 'NO'. There are other such instructions that are not required for malware analysis. So our intermediate language hide/ignore these instructions and make the language transparent to the static analysis. The instructions \emph{MOVD} and \emph{MOVQ} copy a double word or a quad word respectively, from the source operand to the destination operand. Here we have to ask a question do we need to take into account the size of the word being copied in our static analysis? If the answer is 'NO', then in our intermediate language we can replace these kind of instructions with a much simpler \emph{Assignment} instruction. Using such techniques an intermediate language allows us to use simple instructions to make our static analysis much simpler.
\item
We want a common intermediate language that can be used with different platforms, such as Intel x86-64 and ARM. So we do not have to perform separate static analysis for each platform. The intermediate language could be used for any of the above mentioned or other such platforms.
\item
Assembly instructions can have multiple hidden side effects, such as effecting the flags etc, that can substantially increase the efforts required for the static analysis. In this case there are three options that an intermediate language can use to make the static analysis easier: Either remove all the side effects, or have only one side effect, or explicitly define side effect(s) in the instruction. Because our focus is mainly on malware analysis, out of these three, in our opinion the first option is the best option. We will use this option in our intermediate language, and the instructions used in our language that we call MAIL (Malware Analysis Intermediate Language) will not have any side effects.
\end{itemize}

\begin{itemize}

\item
Unknown branch addresses in an assembly makes it difficult to build a correct CFG. This problem will be taken care of by the MAIL. For example, for indirect jumps and calls (branches whose target is unknown or cannot be determined by static analysis) only a change in the source code can change them, so it is safe to ignore these branches for malware analysis where the change is only carried out in the machine code. We explain this using an example from one of the PARSEC \cite{PARSEC} benchmarks.

The following example shows the function \emph{Condition()} from one of the benchmarks in the PARSEC benchmark suite \cite{PARSEC}. This function initializes a static condition variable of a thread. A local variable \emph{rv} is used in a \emph{switch} statement to jump to an appropriate exception generated by a \emph{pthread\_cond\_init()} function. This function initializes the condition variable of a thread and returns zero if successful otherwise returns an error number. The value returned by the \emph{pthread\_cond\_init()} function can only be determined at runtime and so the value of \emph{rv}.

{
\small
\begin{verbatim}
The C++ source code with the translated (disassembled) assembly code:

Condition::Condition(Mutex &_M)
           throw(CondException)
{                                       471b50: push %rbp
   int rv;                              471b51: push %rbx
   M = $_M;                             471b52: sub $0x38,%rsp
   nWaiting = 0;                        471b52: sub $0x38,%rsp
   nWakeupTickets = 0;                  471b56: mov %rsi,(%rdi)
   rv = pthread_cond_init(&c, NULL);    471b59: movl $0x0,0x8(%rdi)
                                        471b60: movl $0x0,0xc(%rdi)
                                        471b67: xor %esi,%esi
                                        471b69: add $0x10,%rdi
                                        471b6d: callq 404b60 <pthread_cond_init@plt>

   switch(rv) {  [  rv UNKNOWN  ]       471b72: cmp $0x16,%eax
      case 0:                           471b75: jbe 471bb0 <Condition:Mutex>
         break;                         471b77: mov 0x21934a(%rip),%r8
      case EAGAIN:                      471b7e: mov $0x8,%edi
      case ENOMEM: {                    471b83: lea 0x10(%r8),%rbp
         CondResourceException e;       471b87: mov %rbp,(%rsp)
         throw e;                       471b8b: callq 404d00 <allocate_exception@plt>
         break;                         471b90: mov 0x219359(%rip),%rdx
      }                                 471b97: mov 0x219342(%rip),%rsi
      case EBUSY:                       471b9e: mov %rax,%rdi
      case EINVAL: {                    471ba1: mov %rbp,(%rax)
         CondInitException e;           471ba4: callq 404da0 <cxa_throw@plt>
         throw e;                       471ba9: nopl 0x0(%rax)
         break;                         471bb0: lea 0x6995(%rip),%rcx <MutexInitException>
      }                                 471bb7: mov %eax,%ebx
      default: {                        471bb9: movslq (%rcx,%rbx,4),%rax
         CondUnknownException e;        471bbd: lea (%rax,%rcx,1),%rdx
         throw e;                       471bc1: jmpq *%rdx     [  UNKNOWN BRANCH TARGET  ]
         break;                         471bc3: nopl 0x0(%rax,%rax,1)
      }                                 471bc8: mov 0x219231(%rip),%rdi
   }                                    471bcf: lea 0x10(%rdi),%rbx
}                                       471bd3: mov $0x8,%edi
                                        471bd8: mov %rbx,0x10(%rsp)
\end{verbatim}
}

Dynamic analysis can be used to determine the value of \emph{rv}, but it is possible that such an analysis may not be able to reach (in one of the runs) one of the executable paths (the \emph{switch} statement) in the case of the \emph{rv} being always zero and changes only in rare cases. These rare cases may not get executed or execute only after running the program for a very long time, that may render the analysis impractical. A malware writer can exploit this weakness and inject the malware code by changing the target address of any of the branches inside the \emph{switch} statement to his/her own malicious code. In such a case the dynamic analysis will not be able to detect this malicious behavior.

That is where static (binary) analysis can help by building a CFG that covers all the available execution paths, in this case the \emph{switch} statement. This CFG may not be correct, because by looking at the disassembled (the assembly) code above we can see it generates an unknown branch target address. This address cannot be computed using static analysis. Is it safe to ignore this branch target address while building the CFG for malware analysis? It is not possible for a malware writer to use this particular instruction as it is for malicious code. He/she will have to change this instruction to make it easy to use, such as the register \emph{rdx} can be loaded with an address of a malicious code before the \emph{jmpq *\%rdx} instruction, which is trivial to detect because in this case the branch target address will become known.

The language MAIL is specifically designed for malware analysis, so we create a new construct/keyword \emph{UNKNOWN} that takes care of these branches. This construct will be helpful not only in static but also in dynamic analysis of the malwares.

\item
A language such as MAIL can be easily translated into a string, a tree or a graph and hence can be optimized for various analysis that are required for malware analysis and detection, such as pattern matching and data mining. Special \emph{patterns} are introduced (Sections \ref{sec:patterns}, \ref{sec:translation} and \ref{sec:subgraph-matching}) in the MAIL language for annotating MAIL statements that can be used for pattern matching.

\item
To reduce the number of different instructions for static analysis, functionally equivalent assembly instructions can be grouped together in one intermediate language instruction, such as:
{
\small
\begin{verbatim}
(xor eax, eax) | (add eax, 0) | (sub eax, eax)  =>  mov eax, 0
(add ebx, 0x2000) & (add eax, ebx) | (lea eax, [ebx + 0x2000])  =>  load eax, expr

where expr = (ebx + 0x2000) and its value can be known or unknown depending on the
value of ebx. This information should be explicitly defined in the language.
\end{verbatim}
}

\end{itemize}

In the following Sections we introduce the new language called MAIL (Malware Analysis Intermediate Language) for malware analysis and detection. In Section \ref{sec:design} we describe its detail design and how a binary program is translated to MAIL. We also cover the CFG (Control Flow Graph) construction and annotation and how graph and pattern matching techniques are used to detect metamorphic malwares \cite{Obfuscation-1}. We carried out an empirical study in Section \ref{sec:study} to test the use of MAIL in a tool. Using the MAIL language the tool was able to fully automate the process of malware analysis and detection and achieved 100\% results. We finaly conclude in Section \ref{sec:conclusion}.

\section{Design of MAIL}\label{sec:design}

In the previous Section we provided motivations for a new language for malware analysis and detection. This Section, introduces and provides the design of, this new language called MAIL. The language MAIL is based on binary analysis to optimize malware detection, so before explaining it's design we first give some background on binary analysis for malware detection.

\subsection{Binary Analysis for Malware Detection}

Almost all the malwares use binaries (instructions that a computer can interpret and execute) to infiltrate a computer system, which can be a desktop, a server, a laptop, a kiosk or a mobile device. Binary analysis is the process of automatically analysing the structure and behavior of a binary program. There are various purposes of this analysis and some of them are: optimization, verification, profiling, performance tuning and computer security. We further explain how binary analysis can help us understand a program and detect malwares in the program, by using a simple binary program (a function called \emph{sort}) that is part of the class \emph{Merge} in a sorting program. This function performs a merge sort on an array of integers. It's binary analysis (performed using an in-house developed tool) information is listed below and explained in the following paragraphs:

{
\scriptsize
\begin{verbatim}
              Listing 1.1 Binary Analysis of The Disassembled Function Merge::sort(int key[], int size)

                        Column I                                                    Column II

0  40108e              55  PUSH                    RBP     :     5  40113b        488b45c8   MOV        RAX, [RBP-0x38]
0  40108f          4889e5   MOV               RBP, RSP     :     5  40113f          4189f9   MOV               R9D, EDI
0  401092              53  PUSH                    RBX     :     5  401142          4189f0   MOV               R8D, ESI
0  401093        4883ec48   SUB              RSP, 0x48     :     5  401145          4889de   MOV               RSI, RBX
0  401097        48897dc8   MOV        [RBP-0x38], RDI     :     5  401148          4889c7   MOV               RDI, RAX
0  40109b        488975c0   MOV        [RBP-0x40], RSI     :     5  40114b      e8e2fdffff  CALL               0x400f32
0  40109f          8955bc   MOV        [RBP-0x44], EDX     :     5  401150          8b45e8   MOV        EAX, [RBP-0x18]
0  4010a2          8b45bc   MOV        EAX, [RBP-0x44]     :     5  401153            01c0   ADD               EAX, EAX
0  4010a5            4898  CDQE                            :     5  401155          0145ec   ADD        [RBP-0x14], EAX
0  4010a7        48c1e002   SHL               RAX, 0x2     :     6  401158          8b45e8   MOV        EAX, [RBP-0x18]
0  4010ab          4889c7   MOV               RDI, RAX     :     6  40115b          8b55bc   MOV        EDX, [RBP-0x44]
0  4010ae      e8e9f9ffff  CALL               0x400a9c     :     6  40115e            89d1   MOV               ECX, EDX
0  4010b3        488945d8   MOV        [RBP-0x28], RAX     :     6  401160            29c1   SUB               ECX, EAX
0  4010b7  c745e801000000   MOV  DWORD [RBP-0x18], 0x1     :     6  401162            89c8   MOV               EAX, ECX
0  4010be      e9f2000000   JMP               0x4011b5 (11):     6  401164          3b45ec   CMP        EAX, [RBP-0x14]
1  4010c3  c745ec00000000   MOV  DWORD [RBP-0x14], 0x0     :     6  401167          0f9fc0  SETG                     AL
1  4010ca      e989000000   JMP               0x401158 (6) :     6  40116a            84c0  TEST                 AL, AL
2  4010cf          8b45e8   MOV        EAX, [RBP-0x18]     :     6  40116c    0f855dffffff   JNZ               0x4010cf (2)
2  4010d2          8b55ec   MOV        EDX, [RBP-0x14]     :     7  401172  c745ec00000000   MOV  DWORD [RBP-0x14], 0x0
2  4010d5          8d0402   LEA         EAX, [RDX+RAX]     :     7  401179      e923000000   JMP               0x4011a1 (9)
2  4010d8          0345e8   ADD        EAX, [RBP-0x18]     :     8  40117e          8b45ec   MOV        EAX, [RBP-0x14]
2  4010db          3b45bc   CMP        EAX, [RBP-0x44]     :     8  401181            4898  CDQE                       
2  4010de    0f8e11000000   JLE               0x4010f5 (4) :     8  401183        48c1e002   SHL               RAX, 0x2
3  4010e4          8b45ec   MOV        EAX, [RBP-0x14]     :     8  401187        480345c0   ADD        RAX, [RBP-0x40]
3  4010e7          8b55bc   MOV        EDX, [RBP-0x44]     :     8  40118b          8b55ec   MOV        EDX, [RBP-0x14]
3  4010ea            89d1   MOV               ECX, EDX     :     8  40118e          4863d2 MOVSXD              RDX, EDX
3  4010ec            29c1   SUB               ECX, EAX     :     8  401191        48c1e202   SHL               RDX, 0x2
3  4010ee            89c8   MOV               EAX, ECX     :     8  401195        480355d8   ADD        RDX, [RBP-0x28]
3  4010f0          2b45e8   SUB        EAX, [RBP-0x18]     :     8  401199            8b12   MOV             EDX, [RDX]
3  4010f3            eb03   JMP               0x4010f8 (5) :     8  40119b            8910   MOV             [RAX], EDX
4  4010f5          8b45e8   MOV        EAX, [RBP-0x18]     :     8  40119d        8345ec01   ADD  DWORD [RBP-0x14], 0x1
5  4010f8          8945e4   MOV        [RBP-0x1c], EAX     :     9  4011a1          8b45ec   MOV        EAX, [RBP-0x14]
5  4010fb          8b45ec   MOV        EAX, [RBP-0x14]     :     9  4011a4          3b45bc   CMP        EAX, [RBP-0x44]
5  4010fe            4898  CDQE                            :     9  4011a7          0f9cc0  SETL                     AL
5  401100        48c1e002   SHL               RAX, 0x2     :     9  4011aa            84c0  TEST                 AL, AL
5  401104          4889c1   MOV               RCX, RAX     :     9  4011ac    0f85ccffffff   JNZ               0x40117e (8)
5  401107        48034dd8   ADD        RCX, [RBP-0x28]     :     10 4011b2          d165e8   SHL  DWORD [RBP-0x18], 0x1
5  40110b          8b45ec   MOV        EAX, [RBP-0x14]     :     11 4011b5          8b45e8   MOV        EAX, [RBP-0x18]
5  40110e          4863d0 MOVSXD              RDX, EAX     :     11 4011b8          3b45bc   CMP        EAX, [RBP-0x44]
5  401111          8b45e8   MOV        EAX, [RBP-0x18]     :     11 4011bb          0f9cc0  SETL                     AL
5  401114            4898  CDQE                            :     11 4011be            84c0  TEST                 AL, AL
5  401116        488d0402   LEA         RAX, [RDX+RAX]     :     11 4011c0    0f85fdfeffff   JNZ               0x4010c3 (1)
5  40111a        48c1e002   SHL               RAX, 0x2     :     12 4011c6        488b45d8   MOV        RAX, [RBP-0x28]
5  40111e          4889c2   MOV               RDX, RAX     :     12 4011ca          4889c7   MOV               RDI, RAX
5  401121        480355c0   ADD        RDX, [RBP-0x40]     :     12 4011cd      e82af9ffff  CALL               0x400afc
5  401125          8b45ec   MOV        EAX, [RBP-0x14]     :     12 4011d2        4883c448   ADD              RSP, 0x48
5  401128            4898  CDQE                            :     12 4011d6              5b   POP                    RBX
5  40112a        48c1e002   SHL               RAX, 0x2     :     12 4011d7              c9 LEAVE                       
5  40112e          4889c3   MOV               RBX, RAX     :     12 4011d8    ff0502000000 INC_A             [RIP+0x02]
5  401131        48035dc0   ADD        RBX, [RBP-0x40]     :     12 4011de            eb04 JMP_A               0x4011e4
5  401135          8b7de4   MOV        EDI, [RBP-0x1c]     :     12 4011e0        00000000 CTR_A                       
5  401138          8b75e8   MOV        ESI, [RBP-0x18]     :     12 4011e4              c3   RET                       

\end{verbatim}
}

The listing shown above is divided into two columns numbered I and II separated by a colon (:). There are total 104 assembly instructions in this function. The first column lists the first 52 instructions and the second column lists the rest of the 52 instructions. This function is part of a binary program (in ELF x86-64 file) that is first disassembled and then binary analysis is performed on the disassembled program for building CFGs of each function in this program. In the listing above each instruction is assigned a block number and an address. Columns I and II are further divided into five columns: Column 1 is the block number, column 2 is the address, column 3 is the machine code, column 4 and 5 are the assembly instructions in Intel syntax.

The total number of blocks in this function are 12. Each block contains different number of instructions. For example block number 4 has only 1 instruction whereas block number 5 has 30 instructions. A block is a basic block \cite{Dragon-Book} that has the following properties: (1) It has only one entry but can have more than one exit points. (2) An instruction with a branch to another block in the same function ends the block. (3) If an instruction is a target of another branch within the same function then that instruction starts a new block.

\begin{figure}[h!t]
\centering
\subfloat[The CFG]{\includegraphics[height=6.0in]{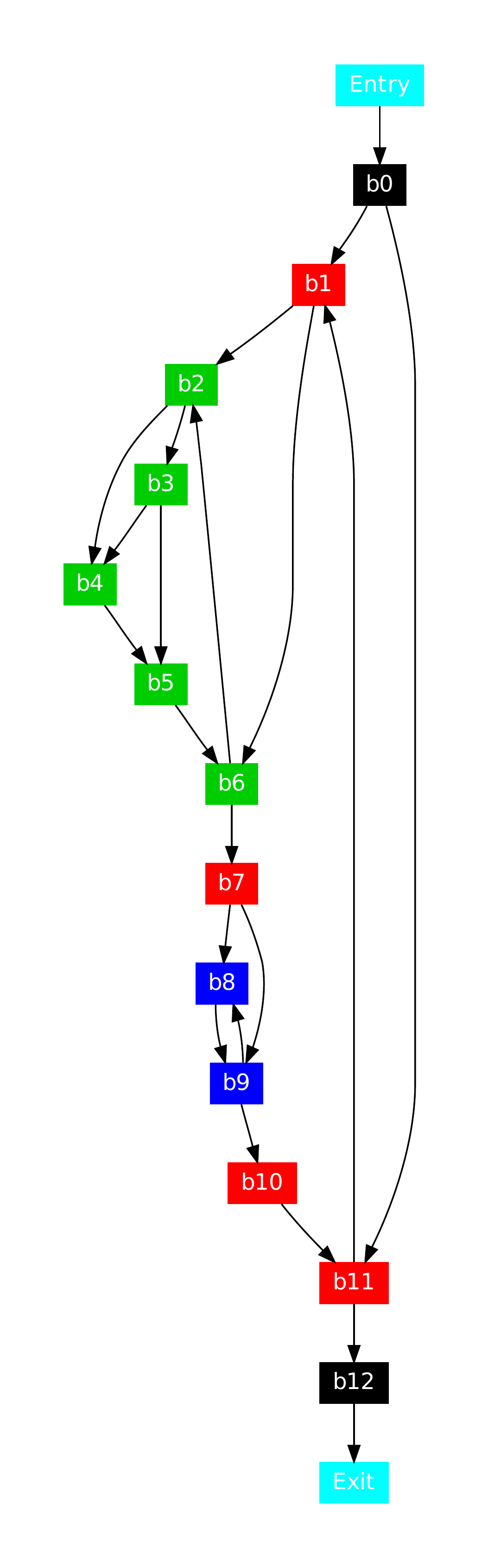}}
\;\;\;\;\;\;\;\;\;\;\;\;\;\;\;\;\;\;\;\;
\subfloat[The Source Code]{\includegraphics[height=4.0in,width=3.3in]{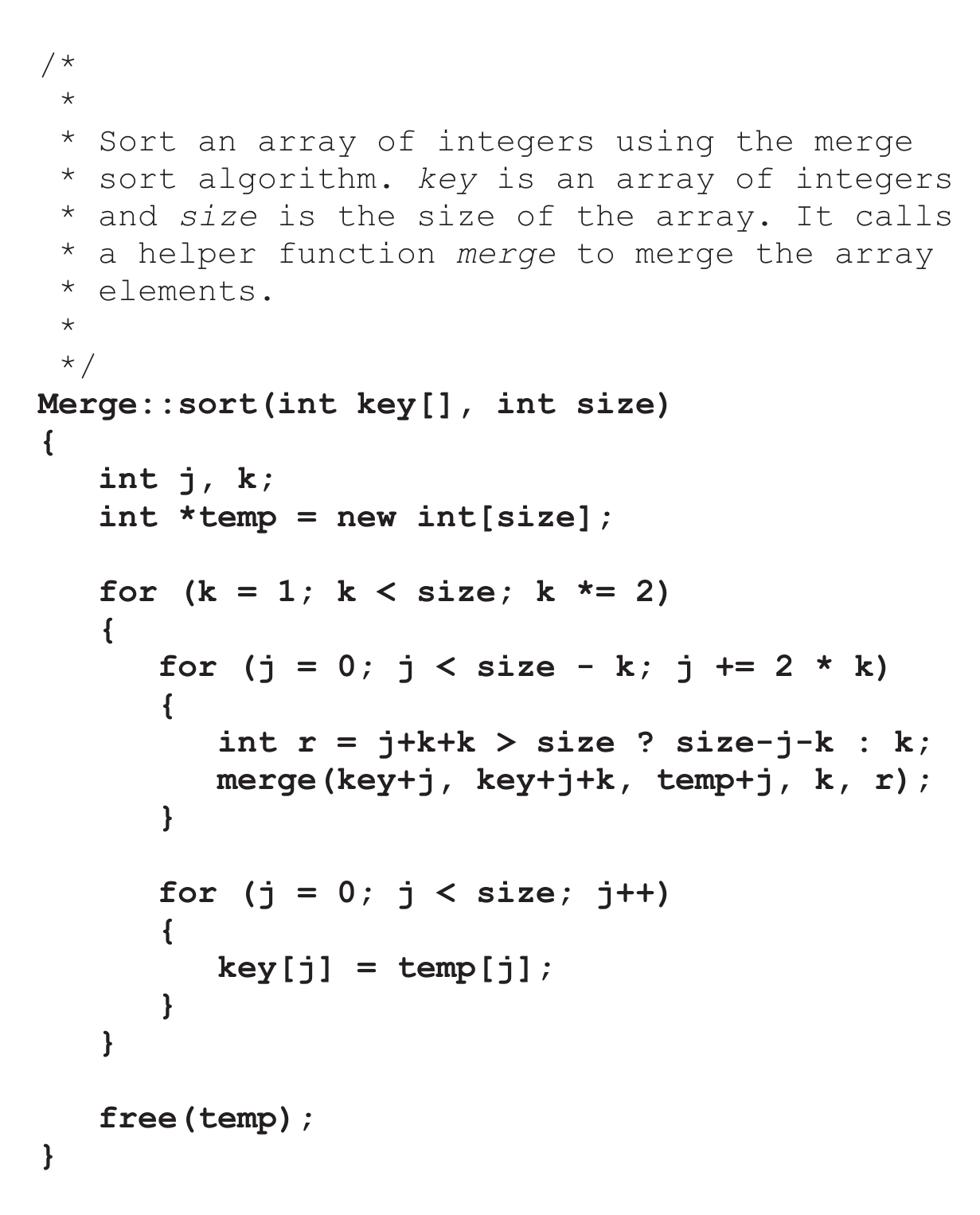}}
\caption{\textbf{The CFG and the Source Code in C++ of the Function in Listing 1.1}}
\label{fig:sample-merge-sort}
\end{figure}

If an instruction branches to another block in the function listed above, the target instruction's block number is listed at the end enclosed in brackets. For example the last instruction in block 1 ends with (6), because this instruction is branching to the address 401158 and the instruction at this address belongs to (is the first instruction of) block 6. Based on the analysis information listed above we build a CFG of this function that is shown in Figure \ref{fig:sample-merge-sort} (a). We are going to compare this CFG with the source code of this function in C++ which is shown in Figure \ref{fig:sample-merge-sort} (b).

The source code was not made available to our binary analysis tool, and the CFG that is build by this tool is only based on the information available in the binary program. Now we want to see how much this analysis helps us to get information about the function for malware detection.

This function has been instrumented, i.e: additional code has been added to this function. There are three instructions at the end (number 101, 102 and 103) in Listing 1.1 that has been added to this function. The first of these instructions INC\_A increment a 32 bit counter CTR\_A at address 4011e0. The second instruction JMP\_A jumps over the counter storage space to address 4011e4 which contains the RET instruction. The counter counts the number of times this function is called. This kind of instrumentation is done for profiling a program for further optimizations.

The CFG starts at block 0 and ends at block 12. Block 11 jumps back to block 1. This indicates a possibility of a loop. If we look at the CFG there is a loop that starts at block 1 and ends at block 11. The blocks that belong to this loop are in red, green and blue colors. This outer loop has two inner loops that are colored green and blue. The source code of the function \emph{Merge::sort()} has exactly one outer loop and the outer loop has exactly two inner loops. \\

A malware writer can change (just the machine code) the last instruction of block 11 in Listing 1.1:

{
\small
\begin{verbatim}
from: 11  4011c0          0f85fdfeffff          JNZ 0x4010c3 (1)
to:   11  4011c0          ebfdfeffffff          JMP 0x4010c3 (1)
\end{verbatim}
}

This will make the outer loop an infinite loop and when the function \emph{Merge::sort()} is called the program will never return. The malware writer in this case has added an unconditional back jump, which in general is a legal jump. Similarly other back jumps (last instructions of blocks 6 and 9) can also be changed by a malware writer to make more infinite loops. So a signature based malware detection tool will not be able to detect such kind of malwares. Without the behavioral information, obtained either statically or dynamically, about a binary program a manual detection with a debugger is required to detect such malwares. This manual labor is very time consuming and financially can become very expensive.

To automate this process, a further binary analysis on the CFG will be able to detect this infinite loop as follows: We have already identified all the identifiable (more on this latter) loops in the function. We will further analyse all the blocks that contain a back jump. In this case analysing block 11, we see that the TEST instruction is followed by an unconditional jump instruction (added by the malware writer), which indicates an illegal infinite loop and hence a malware. In the case where a malware writer also replaces all the other four instructions with some other instructions, we will have to find the unconditional jump instruction to block 1 in block 11 (starting and ending blocks of the loop) to detect the malware, which in the presence of the CFG is trivial.

What if this unconditional jump instruction is a legal instruction, i.e: it has not been added by a malware writer and is part of the program? For example event-based programs contain one or more infinite loops. In this case we may need to build specific control flow patterns and compare them with the previous control flow patterns of malwares of such kind.

Other malicious changes, such as the following register renaming and control flow change, in block 2 in Listing 1.1, cannot be detected by a signature based malware detector:

{
\small
\begin{verbatim}
from:
2  4010cf          8b45e8   MOV        EAX, [RBP-0x18]
2  4010d2          8b55ec   MOV        EDX, [RBP-0x14]
2  4010d5          8d0402   LEA         EAX, [RDX+RAX]
2  4010d8          0345e8   ADD        EAX, [RBP-0x18]
2  4010db          3b45bc   CMP        EAX, [RBP-0x44]
2  4010de    0f8e11000000   JLE               0x4010f5
to:
2  4010cf          8b45e8   MOV        EBX, [RBP-0x18]
2  4010d2          8b55ec   MOV        EDX, [RBP-0x14]
2  4010d5          8d0402   LEA         EBX, [RDX+RBX]
2  4010d8          0345e8   ADD        EBX, [RBP-0x18]
2  4010db          3b45bc   CMP        EBX, [RBP-0x44]
2  4010de    0f8e10100000   JLE               0x4011e5     ; Jump to some malicious code
\end{verbatim}
}

In order to detect such anomaly based malwares automatically, we need control flow information as provided by the binary analysis presented in this Section.

Another technique used by malware writers to deceive signature based detectors is to use instructions other than JMP and CALL to change the control flow of a program. We show this by replacing the last instruction with two instructions in block 7 in Listing 1.1 as follows:

{
\small
\begin{verbatim}
from: 7  401179      e923000000   JMP               0x4011a1
to:   7  401179      68e5114000  PUSH         QWORD 0x4011e5
      7  40117e              c3   RET
\end{verbatim}
}

This change in Listing 1.1 is not finished here. For the code to work correctly the addresses following these instructions and all the effected jump target addresses needs to be updated. A malware writer may or may not update them depending on the complexity of the malware. A tool could be used by the malware writer that can automate updating these addresses.

A further binary analysis on the above instructions reveals that the last value pushed on the stack before the RET statement is 4011e5, so the RET instruction will move the value 4011e5 to the RIP register, the instruction pointer. Next time the instruction at address 4011e5 (malicious code) will be executed.

The added instruction at address 40117e indicates the end of a function. Sometimes the binary provides information about the start and end of all the functions in a program. But if this information is not available it is difficult to find the exact start and end of some of the functions, e.g: the addition of the two instructions shown above in the function in Listing 1.1 divides the function into two functions and makes it difficult to find the original function. For malware detection, we may only need to find where the control is flowing (i.e: just the behavior and not the function boundaries) and then compare this behavior with the previous samples of malwares available to detect such malwares. \\

In the above paragraphs we have shown using an elaborate example, how trivial changes in the binary can make the malware analysis and detection intricate, difficult and expensive. But with suitable tools and appropriate binary analysis it is possible to analyse and detect such malwares automatically. We have build a CFG (Figure \ref{fig:sample-merge-sort} (a)) from the disassembled instructions of the function in Listing 1.1 for malware analysis and detection. In the next Section we describe the design of the intermediate language MAIL that automates and optimizes this step.

\subsection{Design}

We believe a good language must start small and simple, and must give opportunities to the language developers to grow (extend) the language with the users. Therefore MAIL is designed as a small and simple, and an extensible language. In this and next Sections we describe how MAIL is designed in detail.

The basic purpose of the language MAIL is to represent structural and behavioral information of an assembly program for malware analysis and detection. MAIL will also make the program more readable and understandable by a human malware analyst. An assembly program can comprise of the following type of instructions. We use Intel x86-64 assembly instructions \cite{Intel-Developer-Manual} as sample instructions:

\begin{enumerate}
\item
\textbf{Control instructions}: These instructions include instructions that can change the control flow of the program, such as JMP, CALL, RET, CMP, CMPS, CMPPS, PCMPEQW, REP and LOOP instructions.
\item
\textbf{Arithmetic instructions}: These instructions perform arithmetic operations, such as ADD, SUB, MUL, DIV, FSIN, FCOS, PADDW, PSUBW, ADDPS, ADDPD, PMULLD, PAVGW, DPPD, SHR and SHL.
\item
\textbf{Logical instructions}: These include instructions that perform logical operations, such as AND, OR and NOT.
\item
\textbf{Data transfer instructions}: These instructions involve data moving instructions, such as MOV, CMOV, XCHG, PUSH, POP, LODS, STOS, MOVS, MOVAPS, MOVAPD, IN, OUT, INS, OUTS, LAHF, SAHF, PREFETCH, FLDPI, FLDCW, FXSAVE, LEA and LDS.
\item
\textbf{System instructions}: These instructions provide support for operating systems and include instructions LOCK, LGDT, SGDT, LTR, STR and XSAVE etc.
\item
\textbf{Miscellaneous instructions}: All other instructions that do not fit into any other group are included in this group of instructions, such as NOP, CPUID, SCAS, CLC, STC, CLI, HLT, WAIT, MFENCE, PACKSSWB, MAXPS, and UD (undefined instruction).
\end{enumerate}

Designing a language that is small and simple, and accurately represent all these instructions for structural and behavioral information is non-trivial. Our goal is to create as few statements as possible in the intermediate language and map as many instructions as possible to these statements. For example we do not translate (i.e: ignore) the following x86 instructions:

{
\small
\begin{verbatim}
   CLFLUSH:   Flush caches
   CLTS:      Clear TLB)
   SMSW:      Restore machine status word
   VERR:      Verify if a segment can be read
   WBINVD:    Writing back and flushing of external caches
   XRSTOR:    Restore processor extended states from memory
   XSAVE:     Save processor extended states from memory
\end{verbatim}
}

The complete list of x86 and ARM instructions that are not translated into the MAIL statements is given in Appendix \ref{app:ignored-instructions}

\subsection{MAIL Statements}

Majority of the assembly instructions are data moving instructions, as shown above. In the following two MAIL assignment statements we cover the data transfer, arithmetic, logical and some of the system instructions. We use EBNF \cite{EBNF} notation to define these statements:

{
\small
\begin{verbatim}
assignment_s        ::= register_s
                        | address_s ;

register_s          ::= register '=' (math_operator)? expr
                        | register '=' (expr)? math_operator expr
                        | register '=' lib_call_s ;

address_s           ::= address '=' (math_operator)? expr
                        | address '=' (expr)? math_operator expr
                        | address '=' lib_call_s ;

expr                ::= register
                        | address
                        | digit+ ;

register            ::= 'eflags'
                        | 'gr_' digit+
                        | 'fr_' digit+
                        | 'sp'
                        | register_name (':' register_name)? ;

register_name       ::= letter+ ['0' - '9']?
                        | 'ZF'
                        | 'CF'
                        | 'PF'
                        | 'SF'
                        | 'OF' ;

address             ::= '[' digit+ ']'
                        | reg_address
                        | 'UNKNOWN' ;
\end{verbatim}
}

Control instructions are very important because they can change the behavior of a program, and they can be changed or added by polymorphic and metamorphic malwares to avoid detection. The following MAIL control statement represent the control instructions:

{
\small
\begin{verbatim}
control_s           ::= ( 'if' condition_s (jump_s | assignment_s) )
                        ( 'else' (jump_s | assignment_s) )? ;

jump_s              ::= 'jmp' address ;

lib_call_s          ::= letter+ '(' address (, args)* ')' ;

function_s          ::= 'start_function_' digit+ statement 'end_function_' digit+ ;

condition_s         ::= (expr rel_operator expr)+ ;
\end{verbatim}
}

All the MAIL language statements can be divided into the following 8 basic statements. The complete MAIL grammar is given in Appendix \ref{app:grammar-mail}:

{
\small
\begin{verbatim}
statements          ::= ( statement* ) ;
statement           ::= assignment_s+
                        | control_s+
                        | condition_s+
                        | function_s+
                        | jump_s+
                        | lib_call_s+
                        | 'halt'
                        | 'lock' ;
\end{verbatim}
}

Every statement in the MAIL language has a \emph{type} also called a \emph{pattern} that can be used for pattern matching during malware analysis and detection. These \emph{patterns} are introduced and explained in Section \ref{sec:patterns}. MAIL has its own registers but also reuses the registers present in the architecture that is being translated to the MAIL language. There are other special registers such as:

\textbf{Flag registers:} ZF (zero flag), CF (crry flag), PF (parity flag), SF (sign flag) and OF (overflow falg). These flag registers are of size one byte and are used in conditional statements. e.g: if (ZF == 1) jmp 405632;. \textbf{eflags:} stores the flag registers. \textbf{sp:} To keep track of the stack pointer. \textbf{gr and fr:} These are infinite number of general purpose registers for use in integer and floating point instructions respectively.

\subsection{MAIL Library}\label{sec:mail-library}

We have added 22 library functions to the MAIL language. Table \ref{tab:mail-library} gives details about all these library functions. These library functions can help in translating most of the complex assembly instructions present in current processors architecture. The purpose of these functions is not to capture the exact functionality of the assembly instruction(s) but to help in analysing the structure and the behavior of the assembly program, and capturing some of the patterns in the program that can help detect malwares.

\begin{table}[h!]
\centering
\small
{
\begin{tabular}{  l  l  } \hline\noalign{\smallskip}

 \textbf{Function}          & \textbf{Semantics}                                                                            \\ \noalign{\smallskip}\hline\hline\noalign{\smallskip}
 abs(op)                    & Returns the absolute value of the parameter \emph{op}                                         \\ \noalign{\smallskip}\hline\noalign{\smallskip}
 aes(op, mode)              & Performs AES encryption/decryption on \emph{op}; mode=0 for encrypt and vice versa            \\ \noalign{\smallskip}\hline\noalign{\smallskip}
 allocate(n)                & Allocate memory from the heap of size \emph{n} bytes                                          \\ \noalign{\smallskip}\hline\noalign{\smallskip}
 atan(op)                   & Returns the arc tangent of the parameter \emph{op}                                            \\ \noalign{\smallskip}\hline\noalign{\smallskip}
 avg(op1, op2)              & Computes the average of the parameters \emph{op1} and \emph{op2}                              \\ \noalign{\smallskip}\hline\noalign{\smallskip}
 bit(op, index, len)        & Selects \emph{len} number of bits in \emph{op} starting at \emph{index}                       \\ \noalign{\smallskip}\hline\noalign{\smallskip}
 clear(op, index, len)      & Clears the bits in \emph{op} at \emph{index} upto \emph{len}                                  \\ \noalign{\smallskip}\hline\noalign{\smallskip}
 compare(op1, op2)          & Compares two values \emph{op1} and \emph{op2} and then set the flag register                  \\ \noalign{\smallskip}\hline\noalign{\smallskip}
 complement(op, index)      & Complements the bit in \emph{op} at \emph{index}                                              \\ \noalign{\smallskip}\hline\noalign{\smallskip}
 convert(value)             & Convert the \emph{value} to either int or float                                               \\ \noalign{\smallskip}\hline\noalign{\smallskip}
 cos(op)                    & Returns the cosine of the parameter \emph{op}                                                 \\ \noalign{\smallskip}\hline\noalign{\smallskip}
 count(op)                  & Counts the number of ones in the \emph{op}                                                    \\ \noalign{\smallskip}\hline\noalign{\smallskip}
 len(obj)                   & Computes the length of the parameter \emph{obj}                                               \\ \noalign{\smallskip}\hline\noalign{\smallskip}
 log(op)                    & Computes the log of the parameter \emph{op}                                                   \\ \noalign{\smallskip}\hline\noalign{\smallskip}
 max(op1, op2)              & Returns the maximum of the parameters \emph{op1} and \emph{op2}                               \\ \noalign{\smallskip}\hline\noalign{\smallskip}
 min(op1, op2)              & Returns the minimum of the parameters \emph{op1} and \emph{op2}                               \\ \noalign{\smallskip}\hline\noalign{\smallskip}
 rev(op)                    & Reverses the bit order in op                                                                  \\ \noalign{\smallskip}\hline\noalign{\smallskip}
 round(op)                  & Rounds the parameter \emph{op}                                                                \\ \noalign{\smallskip}\hline\noalign{\smallskip}
 scanf(op1, op2)            & Stores the index of the first bit one, found in \emph{op1}, in \emph{op2} (forward scan)      \\ \noalign{\smallskip}\hline\noalign{\smallskip}
 scanr(op1, op2)            & Stores the index of the last bit one, found in \emph{op1}, in \emph{op2} (reverse scan)       \\ \noalign{\smallskip}\hline\noalign{\smallskip}
 set(op, index, len)        & Sets the bit in \emph{op} at \emph{index} upto len                                            \\ \noalign{\smallskip}\hline\noalign{\smallskip}
 sin(op)                    & Returns the sine of the parameter \emph{op}                                                   \\ \noalign{\smallskip}\hline\noalign{\smallskip}
 sqrt(op)                   & Computes the square root of the parameter \emph{op}                                           \\ \noalign{\smallskip}\hline\noalign{\smallskip}
 substr(value, offset, len) & Returns the sub string from the string \emph{value} starting at \emph{offset} upto \emph{len} \\ \noalign{\smallskip}\hline\noalign{\smallskip}
 swap(op1, op2)             & Swaps the bits in \emph{op2} and write back in \emph{op1}                                     \\ \noalign{\smallskip}\hline\noalign{\smallskip}
 swap(op)                   & Swaps the bits in \emph{op}                                                                   \\ \noalign{\smallskip}\hline\noalign{\smallskip}
 tan(op)                    & Returns the tangent of the parameter \emph{op}                                                \\ \noalign{\smallskip}\hline\noalign{\smallskip}

\end{tabular}
\\ [5pt]
}
\caption{\textbf{MAIL Library Functions}}
\label{tab:mail-library}
\end{table}

\subsection{MAIL Patterns for Annotation}\label{sec:patterns}

MAIL language can also be used to annotate a CFG of a program using different patterns available in the MAIL language. The purpose of these annotations is to assign patterns to MAIL statements that can be used latter for pattern matching during malware detection. Section \ref{sec:translation} gives an example of a CFG with pattern annotation and Section \ref{sec:subgraph-matching} explains how they are used in malware detection. More than one statements in the MAIL langauge can have one pattern. There are total 21 patterns in the MAIL language and are listed and explained as follows:

\begin{itemize}
\item
\textbf{ASSIGN:} An assignment statement. \emph{e.g:} EAX = EAX + ECX;
\item
\textbf{ASSIGN\_CONSTANT:} An assignment statement including a constant. \emph{e.g:} EAX = EAX + 0x01;
\item
\textbf{CONTROL:} A control statement where the target of the jump is unknown. \emph{e.g:} if (ZF == 1) JMP [EAX+ECX+0x10];
\item
\textbf{CONTROL\_CONSTANT:} A control statement where the target of the jump is known. \emph{e.g:} if (ZF == 1) JMP 0x400567;
\item
\textbf{CALL:} A call statement where the target of the call is unknown. \emph{e.g:} CALL EBX;
\item
\textbf{CALL\_CONSTANT:} A call statement where the target of the call is known. \emph{e.g:} CALL 0x603248;
\item
\textbf{FLAG:} A statement including a flag. \emph{e.g:} CF = 1;
\item
\textbf{FLAG\_STACK:} A statement including flag register with stack. \emph{e.g:} EFLAGS = [SP=SP-0x1];
\item
\textbf{HALT:} A halt statement. \emph{e.g:} HALT;
\item
\textbf{JUMP:} A jump statement where the target of the jump is unknown. \emph{e.g:} JMP [EAX+ECX+0x10];
\item
\textbf{JUMP\_CONSTANT:} A jump statement where the target of the jump is known. \emph{e.g:} JMP 0x680376
\item
\textbf{JUMP\_STACK:} A return jump. \emph{e.g:} JMP [SP=SP-0x8]
\item
\textbf{LIBCALL:} A library call. \emph{e.g:} compare(EAX, ECX);
\item
\textbf{LIBCALL\_CONSTANT:} A library call including a constant. \emph{e.g:} compare(EAX, 0x10);
\item
\textbf{LOCK:} A lock statement. \emph{e.g:} lock;
\item
\textbf{STACK:} A stack statement. \emph{e.g:} EAX = [SP=SP-0x1];
\item
\textbf{STACK\_CONSTANT:} A stack statement including a constant. \emph{e.g:} [SP=SP+0x1] = 0x432516;
\item
\textbf{TEST:} A test statement. \emph{e.g:} EAX and ECX;
\item
\textbf{TEST\_CONSTANT:} A test statement including a constant. \emph{e.g:} EAX and 0x10;
\item
\textbf{UNKNOWN:} Any unknown assembly instruction that cannot be translated.
\item
\textbf{NOTDEFINED:} The default pattern. \emph{e.g:} All the new statements when created are assigned this default value.
\end{itemize}

\subsection{Binary to MAIL Translation}\label{sec:translation}

\begin{figure}[h!t]
\centering
\scalebox{0.80}
{\includegraphics{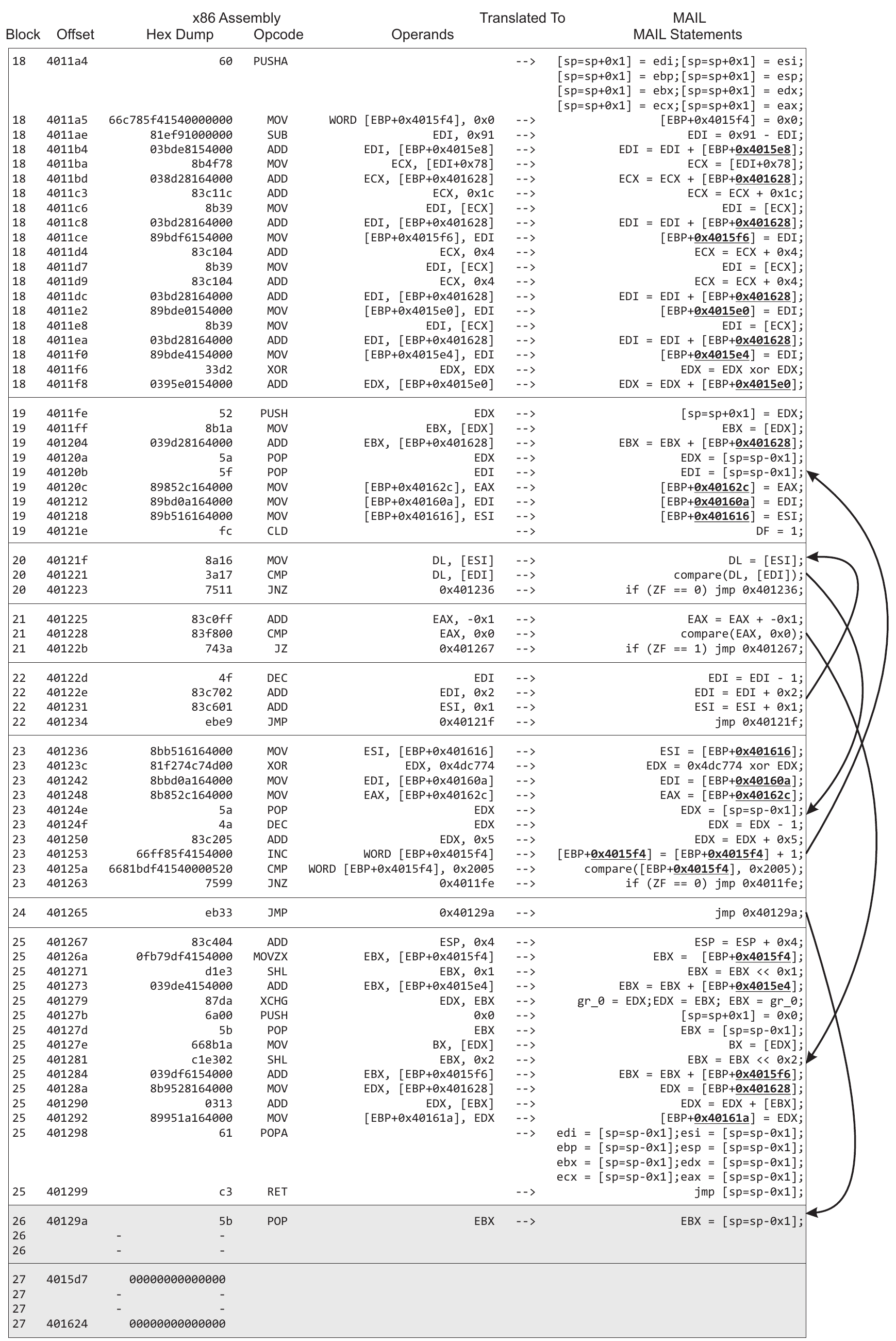}}
\caption{\textbf{Disassembly and Translation to MAIL of one of the Functions of one of the Malware Samples Listed in Table \ref{tab:data-set}}}
\label{fig:malware-sample}
\end{figure}

For translating a binary program, we first disassemble the binary program into an assembly program. Then we translate this assembly into MAIL program. We use a sample program, one of the malware samples, to give an example of the steps invlove in the translation as shown in Figure \ref{fig:malware-sample}. This example shows how x86 assembly program is translated to MAIL program. In Section \ref{sec:arm-assembly-translation} we give an example of translating an ARM assembly program to MAIL program.

The binary analysis of the function shown in Figure \ref{fig:malware-sample} have identified 5 blocks in this function labelled 18 - 25. There are two columns separated by $\rightarrow$. The first column lists the x86 assembly instructions and the second column lists the corresponding translated MAIL statements. The \emph{mathematical} instructions are translated to an assignment statement with the appropiate operator added. Most of the \emph{data} instructions are translated to simple assignement statements. Conditional \emph{jump} instructions, such as \emph{JZ} and \emph{JNZ}, are translated to an \emph{if} statement. Some of the instructions are translated to more than one MAIL statements. For example the instruction \emph{XCHG} in block 25 is translated to three MAIL statements. The MAIL library functions are used to translate some of the instructions, such as the instrucion \emph{CMP} in blocks 20 and 21 is translated using the library function \emph{compare}. All the MAIL library functions are explained in Section \ref{sec:mail-library}

In addition to its own registers the MAIL language reuses all the x86 registers. There is a special register \emph{sp} used in the MAIL language to keep track of the stack pointer in the program. The example shows data embedded inside the \emph{code section} in block 27. This block is used to store, load and process data by the function as pointed out by the underlined addresses in the blocks. There are five instructions that change the control flow of the program and are indicted by the arrows in the Figure. There are two back edges $20 \leftarrow 22$ and $19 \leftarrow 23$. These edges indicate the presence of loops in the program. The \emph{jump} in block 24 jumps out of the function. MAIL also keeps track of the flags using boolean values. For example the instruction \emph{CLD} sets the direction flag in block 19.

Each MAIL statement is associated with a \emph{type} also called a \emph{pattern}. There are total 21 \emph{patterns} in the MAIL language as explained in Section \ref{sec:patterns}. For example an assignment statement with a constant value and an assignment statement without a constant value are two different \emph{patterns}. Jump statements can have upto three \emph{patterns}. Following are the \emph{patterns} that are assigned during translation to the statements in block 21 shown in Figure \ref{fig:malware-sample}:

{
\small
\begin{verbatim}
 21               EAX = EAX + -0x1;     -->     ASSIGN_CONSTANT
 21              compare(EAX, 0x0);     -->     CALL_CONSTANT
 21      if (ZF == 1) jmp 0x401267;     -->     CONTROL_CONSTANT

\end{verbatim}
}

These patterns are used to annotate a CFG for pattern matching. Section \ref{sec:subgraph-matching} explains how they help and improve malware detection.

\subsubsection{Translation of an ARM Program to MAIL Program}\label{sec:arm-assembly-translation}

\begin{figure}[h!t]
\centering
\scalebox{0.70}
{\includegraphics{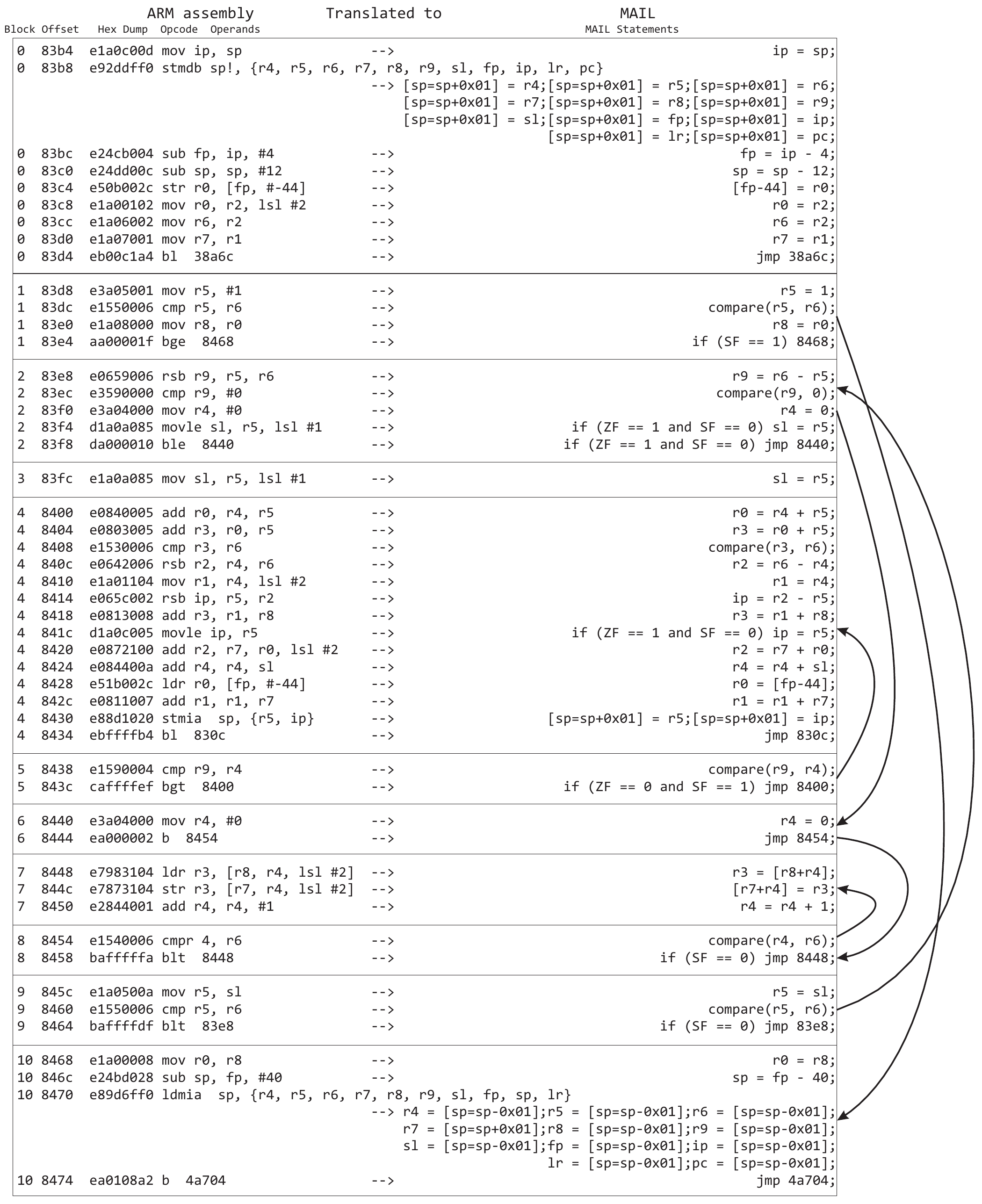}}
\caption{\textbf{An Example of Translation of an ARM Assembly Program to MAIL Program. The Function Merge::sort() shown in Figure \ref{fig:sample-merge-sort} is First Disassembled to ARM Assembly Program and Then Translated to MAIL Program.}}
\label{fig:sample-arm-translation}
\end{figure}

In the previous Section we gave an example of translating x86 assembly program to MAIL program. As already mentioned that MAIL as a common language for different platforms, such as x86 and ARM binaries, helps malware analysis and detection tools to achieve platform independence. In this Section we provide an example of translating an ARM assembly program to MAIL program. Figure \ref{fig:sample-arm-translation} shows an example of such a translation.

\begin{figure}[h!t]
\centering
\scalebox{0.50}
{\includegraphics{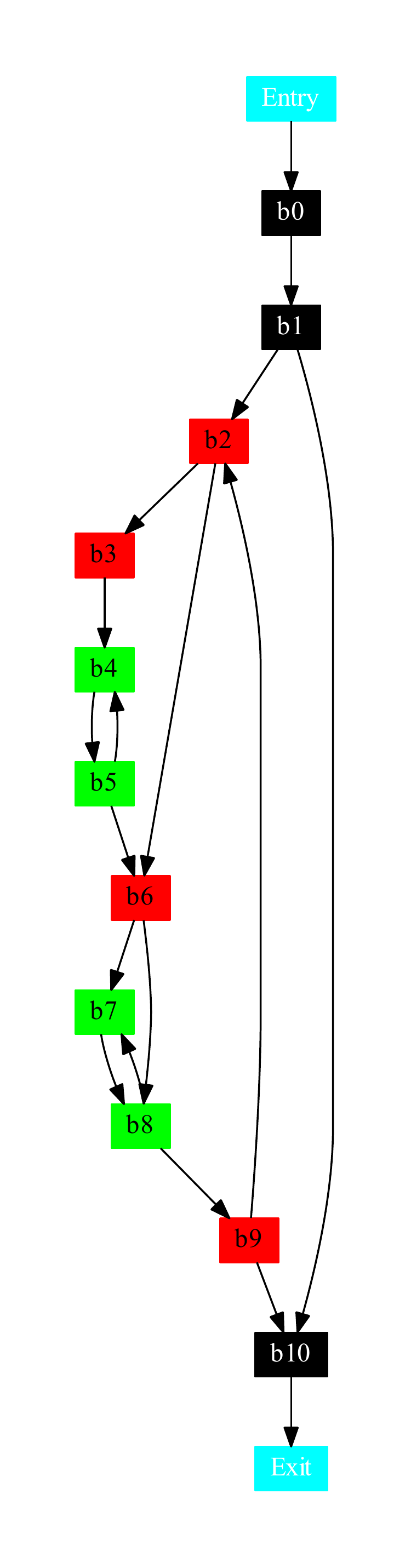}}
\caption{\textbf{CFG of the Function Shown in Figure \ref{fig:sample-arm-translation}. Except One of the Inner Loops this CFG is almost similar to the CFG shown in Figure \ref{fig:sample-merge-sort} (a). The CFG Shown in Figure \ref{fig:sample-merge-sort} (a) is of the Same Function But Constructed From the x86 Program.}}
\label{fig:cfg-sample-arm-translation}
\end{figure}

The main obvious difference in an ARM program and a x86 program is the length of the instructions as shown in Figures \ref{fig:malware-sample} and \ref{fig:sample-arm-translation}. The size of the hex dump of each instruction of the x86 program is different. Whereas the size of the hex dump of the ARM program is the same for each instruction. For the sake of completeness we also show the CFG of this program in Figure \ref{fig:cfg-sample-arm-translation}.

As shown in Figure \ref{fig:sample-arm-translation} the binary analysis have identified 10 blocks in the program. The \emph{mathematical}, \emph{load} and \emph{mov} instructions are translated to assignment statements. The one conditional move \emph{movele} instruction is translated to a \emph{control} statement. Branch instructions are translated to either simple \emph{jump} or \emph{control} statements. The instruction \emph{cmp} is translated using the MAIL library function \emph{compare()}.

\subsection{CFG Construction}

\begin{figure}[h!t]
\centering
\scalebox{0.50}
{\includegraphics{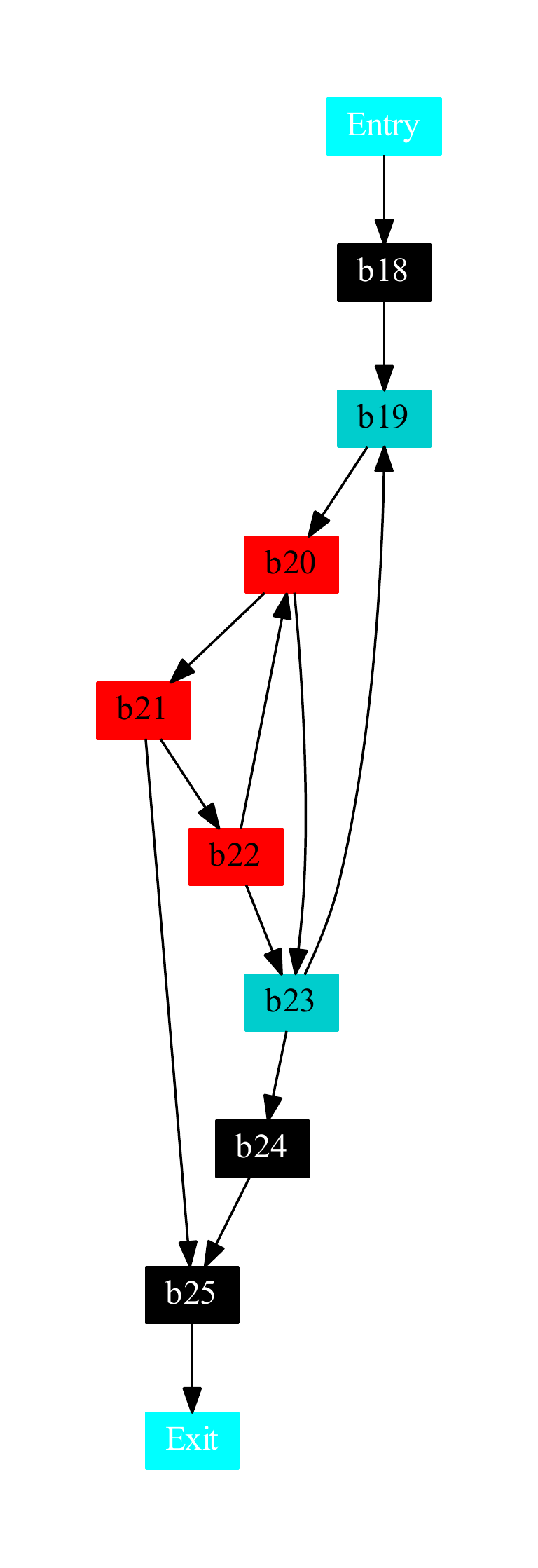}}
\caption{\textbf{CFG of the Function Shown in Figure \ref{fig:malware-sample}}}
\label{fig:cfg-malware-sample}
\end{figure}

Figure \ref{fig:cfg-malware-sample} shows the CFG of the sample program shown in Figure \ref{fig:malware-sample}. The CFG clearly indicates two back edges and two forward edges that change the control flow of the program. The fifth edge that jumps out of the function shown in Figure \ref{fig:malware-sample} is not shown in this CFG. There are two loops one outer loop \{19, 20, 21, 22, 23\} and one inner loop \{20, 21, 22\}.

\subsection{Subgraph and Pattern Matching}\label{sec:subgraph-matching}

After the binary analysis performed above we get a CFG of a program as shown in Figure \ref{fig:cfg-malware-sample}. For detecting if a program contains a malware we compare the CFG of a program with the CFG of a known malware. If the CFG of the malware matches the complete or part of the CFG of the program then the program contains a malware, i.e; the program is not benign. We formulate this problem of malware detection as follows:

Let $G = (V,E)$ is graph of the program and $M = (V',E')$ is graph of the malware, where $V, V'$ and $E, E'$ are the vertices and edges of the graphs respectively. Let $G_{sg} = (V_{sg}, E_{sg})$ where $V_{sg} \subseteq V$ and $E_{sg} \subseteq E$. If $G_{sg} \cong M$ then $G$ is not benign.

We solve this problem using subgraph isomorphism (matching). Given the input of two graphs it determines if one of the graphs contains a subgraph that is isomorphic (similar in shape) to the other graph. Generally subgraph isomorphism is an NP-Complete problem \cite{Cook-Isomorphism-NP-Complete}. A CFG of a program is a sparse graph therefore it is possible to compute the isomorphism of two CFGs in a reasonable amount of time.

Very small graphs when matched against a large graph can produce a false positive matching. We conducted an experiment (more details in the next paragraph) and found that some of the malware samples after normalization were reduced to a small graph of 3 nodes as shown in Table \ref{tab:data-set} and were responsible for producing a large number (87.85\%, see Table \ref{tab:sg-pm-results}) of false positives. To take care of these and other such graphs we also implemented a \emph{Pattern Matching} sub-component within the \emph{Subgraph Matching} component. Every statement in the language MAIL is assigned a \emph{pattern} as explained in Section \ref{sec:patterns}. We use this \emph{pattern} to match each statement in the matching nodes of the two graphs.

\begin{figure}[h!t]
\centering
{\includegraphics[scale=0.80]{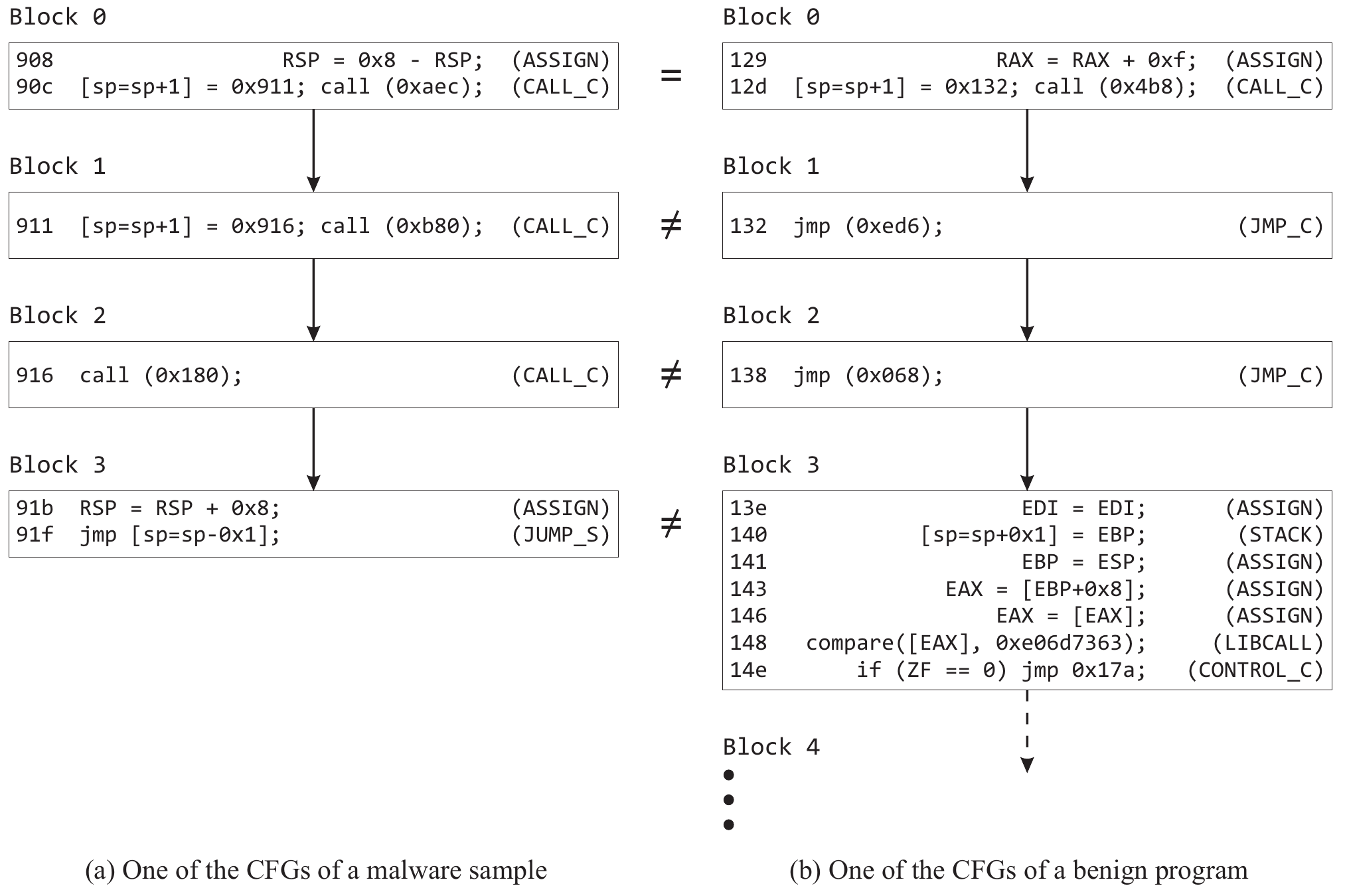}}
\caption{\textbf{Example of \emph{pattern matching} of two \emph{isomorphic} CFGs. The CFG in (a) is \emph{isomorphic} to the subgraph (blocks 0 - 3) of the CFG in (b).}}
\label{fig:pattern-matching-example}
\end{figure}

An example of \emph{Pattern Matching} of two \emph{isomorphic} CFGs is shown in Figure \ref{fig:pattern-matching-example}. One of the CFGs of a malware sample, shown in Figure \ref{fig:pattern-matching-example} (a), is \emph{isomorphic} to a subgraph of one of the CFGs of a benign program, shown in Figure \ref{fig:pattern-matching-example} (b). Considering these two CFGs as a match for malware detection will produce a wrong result, a false positive. The statements in the benign program do not match with the statements in the malware sample. To reduce this false positive we have two options: (1) we can match each statement exactly with each other or (2) assign patterns to these statements for matching. Option (1) will not be able to detect unknown malware samples and is time consuming, so we use option (2) in our approach, which in addition to reducing false positives has the potential of detecting unknown malware samples.

For a successful \emph{pattern matching} we require all the statements in the matching \emph{blocks} to have the same patterns. In Figure \ref{fig:pattern-matching-example}, only the statements in \emph{block 0} satisfy this requirement. The statements in all the other blocks do not satisfy this requirement, therefore these CFGs fail the \emph{pattern matching}.

To verfiy that by adding the \emph{Pattern Matching} component to the \emph{Subgraph Matching} component we have improved the metamorphic malware matching technique, \textbf{an experimental study} was performed. The dataset used for the experiment consisted of total 4289 samples, including 266 malware samples. Out of the 266 malware samples 250 were metamorphic malwares. Out of these 250 metamorphic malware samples we randomly selected 27 samples (10.11\% of all the malware samples) for use in the experiment for matching. Dataset distribution based on the size of the CFG after normalization is shown in Table \ref{tab:data-set-small-study}.

\setlength{\tabcolsep}{20pt}
\begin{table}[h!]
\centering
{
\begin{tabular}{  c  c  c  c  } \hline\noalign{\smallskip}

\multicolumn{2}{ c }{27}            & \multicolumn{2}{ c }{4289}               \\
\multicolumn{2}{ c }{Malware Samples Used For Matching}& \multicolumn{2}{ c }{Benign and Malware Samples}    \\ \noalign{\smallskip}
 \textbf{Size of} & \textbf{Number of} & \textbf{Size of} & \textbf{Number of}  \\ 
 \textbf{CFG}     & \textbf{Samples}   & \textbf{CFG}     & \textbf{Samples}    \\ \noalign{\smallskip}\hline\hline\noalign{\smallskip}

 3                & 19                 & 0 -- 4           & 476     \\ \noalign{\smallskip}\hline\noalign{\smallskip}
 92               & 1                  & 5 -- 200         & 402      \\ \noalign{\smallskip}\hline\noalign{\smallskip}
 93               & 2                  & 201 -- 1000      & 753     \\ \noalign{\smallskip}\hline\noalign{\smallskip}
 95               & 1                  & 1001 -- 4997     & 1215     \\ \noalign{\smallskip}\hline\noalign{\smallskip}
 96               & 1                  & 5007 -- 9987     & 579      \\ \noalign{\smallskip}\hline\noalign{\smallskip}
 99               & 2                  & 10012 -- 19979   & 464      \\ \noalign{\smallskip}\hline\noalign{\smallskip}
 100              & 1                  & 20032 -- 29889   & 187      \\ \noalign{\smallskip}\hline\noalign{\smallskip}
                  &                    & 30109 -- 63289   & 213      \\ \noalign{\smallskip}\hline\noalign{\smallskip}
 
\end{tabular}
\\ [5pt]
}
\caption{\textbf{Data Set Distribution, For the Experimental Study, Based on the Size (number of nodes) of the Control Flow Graph (CFG) after Normalization}}
\label{tab:data-set-small-study}
\end{table}
\setlength{\tabcolsep}{6pt}

The complexity (size) of these graphs range from 0 nodes to 63289 nodes. Some of the Windows DLLs (dynamic link libraries) that were used in the experiment do not have code but only data (cannot be executed) and that is why they have 0 node graphs (CFGs). All the 4289 samples were matched against the 27 selected malware samples. The results of this experiment are shown in Table \ref{tab:sg-pm-results}. There are 87.85\% false positives when only the \emph{Subgraph Matching} technique is used. The reason for this large number of false positives, as explained earlier, is because of the small size (3 nodes as shown in Table \ref{tab:data-set-small-study}) of the graph samples matched. This large false positive has been reduced to 0 when in addition to the \emph{Subgraph Matching} technique the \emph{Pattern Matching} technique is used.

\textbf{An interesting observation:} When we used the \emph{Pattern Matching} technique in addition to the \emph{Subgraph Matching} technique the number of graphs matched were not 27 but 168. The 27 malware graph samples were not only matched with the same 27 graphs but they were also matched with additional 141 metamorphic malware graphs. When checked manually the nodes matched contained the same patterns as found in one of the 27 malware graph samples used for matching. That means some of the unknown metamorphic malwares were also detected because of the use of the \emph{Pattern Matching} technique.

\begin{table}[h!]
\centering
\footnotesize
{
\begin{tabular}{  c  c  c  c  c  c  } \hline\noalign{\smallskip}

\textbf{Component(s) Used} & \textbf{TNG \footnotemark[1]} & \textbf{NGUM \footnotemark[2]} & \textbf{NGM \footnotemark[3]} & \textbf{NMGM \footnotemark[4]} & \textbf{False Positives} \\ \noalign{\smallskip}\hline\hline\noalign{\smallskip}

\emph{Subgraph Matching} & 4289                  & 27 & 4018/93.68\%              & 257 \footnotemark[5] & 3768/87.85\% \\ \noalign{\smallskip}\hline\noalign{\smallskip}
\emph{Subgraph Matching} &                       &    &                           &                      &  \\
and                      & 4289                  & 27 & 168/3.92\%                & 168 \footnotemark[6] & 0/0\% \\
\emph{Pattern Matching}  &                       &    &                           &                      &  \\ \noalign{\smallskip}\hline\noalign{\smallskip}

\end{tabular}
\\ [5pt]
\scriptsize
{
\begin{tabular}{  l  }
1 \;\;TNG:\;\;\;\;\;Total number of graphs (both benign and malware samples) \\
2 \;\;NGUM:\;\;Number of graphs (only malware samples) used for matching \\
3 \;\;NGM:\;\;\;\;\;Number of graphs (both benign and malware samples) matched \\
4 \;\;NMGM:\;\;Number of malware graphs matched \\
5 \;\;The graphs matched were the 250 metamorphic malwares and the 7 other \\
  \;\;\;\;\;malwares. \\
6 \;\;\textbf{An interesting observation:} The 27 malware graph samples were not \\
  \;\;\;\;\;only matched with the same 27 graphs but they were also matched with \\
  \;\;\;\;\;additional 141 metamorphic malware graphs. When checked manually the \\
  \;\;\;\;\;nodes matched contained the same patterns as found in one of the 27 \\
  \;\;\;\;\;malware graph samples used for matching. \\ \\

Machine used: \;\;\;\;\;\;\;Intel Core i5 CPU M 430 @ 2.27 GHz \\
RAM: \;\;\;\;\;\;\;\;\;\;\;\;\;\;\;\;\;\;\;\;4 GB \\
Operating System: Windows 8 Professional \\
\end{tabular}
}
}
\caption{\textbf{Results of the Experiment to Verify the Improved Matching Technique After Adding the \emph{Pattern Matching} Componenent.}}
\label{tab:sg-pm-results}
\end{table}

This experimental study confirms the improved results of these two matching components for detecting an already known metamorphic malware and the 100\% detection rate we achieved also confirms these results as shown in Table \ref{tab:final-results}.

\section{Empirical Study of Using the MAIL Language}\label{sec:study}

We carried out an empirical study to analyse the correctness and the efficiency of our techniques described above using the MAIL language. This Section describes this empirical study. We developed a prototype tool called MARD that uses MAIL for malware analysis and detection as described above. We collected different metamorphic malwares and Windows programs as samples to use in our tool.

\subsection{Dataset}

Our dataset for the experiments consisted of 1387 programs. Out of these: 250 are metamorphic malware samples collected from two different resources \cite{Histogram-MD, Opcode-Graph-MD}, and the other 1137 are Windows benign programs. Table \ref{tab:data-set} gives more details about this dataset.

\setlength{\tabcolsep}{8pt}
\begin{table}[h!]
\centering
{
\begin{tabular}{  c  c  c  c  } \hline\noalign{\smallskip}

\multicolumn{2}{ c }{250}            & \multicolumn{2}{ c }{1137}               \\
\multicolumn{2}{ c }{Malware Samples}& \multicolumn{2}{ c }{Benign Programs}    \\ \noalign{\smallskip}
 \textbf{Size of} & \textbf{Number of} & \textbf{Size of} & \textbf{Number of}  \\ 
 \textbf{CFG}     & \textbf{Samples}   & \textbf{CFG}     & \textbf{Samples}    \\ \noalign{\smallskip}\hline\hline\noalign{\smallskip}

 3                & 200                & 17               & 127     \\ \noalign{\smallskip}\hline\noalign{\smallskip}
 88               & 1                  & 30               & 44      \\ \noalign{\smallskip}\hline\noalign{\smallskip}
 91 -- 99         & 38                 & 44 -- 998        & 412     \\ \noalign{\smallskip}\hline\noalign{\smallskip}
 100 -- 104       & 10                 & 1000 -- 9765     & 535     \\ \noalign{\smallskip}\hline\noalign{\smallskip}
 129              & 1                  & 10118 -- 15343   & 19      \\ \noalign{\smallskip}\hline\noalign{\smallskip}
 
\end{tabular}
\\ [5pt]
}
\caption{\textbf{Data Set Distribution, For the Empirical Study, Based on the Size (number of nodes) of the Control Flow Graph (CFG) after Normalization}}
\label{tab:data-set}
\end{table}
\setlength{\tabcolsep}{6pt}

The data set contains a variety of programs with simple CFGs to complex CFGs for testing. As shown the size of the CFG of the malware samples range from 3 nodes to 129 nodes, and the size of the CFG of the benign programs range from 17 nodes to 15343 nodes. This variety in the samples provides a good testing platform for graph and pattern matching techniques used in our tool.

\subsection{Experiments and Results}\label{sec:experiments}

Two experiments were carried out using our tool MARD to detect metamorphic malwares: \textbf{(1)} In the first experiment we wanted to see if the tool can detect all the known malwares. This experiment consisted of 250 known malware samples in the test data set. \textbf{(2)} In the second experiment we wanted to see if the tool can detect the unknown malwares. This experiment consisted of 225 unknown malware samples in the test data set. The results for this experiment were obtained using 10-fold cross validation. This Section gives details about these experiments and the results obtained.

\subsubsection{Experiment (1): To detect known malwares}

The tool MARD first builds the training dataset, also called \emph{Malware Templates}, using the 250 malware samples. After a program (sample) is translated to MAIL and to a CFG the tool detects the presence of malwares in the program, using the \emph{Malware Templates} and applying the graph and the pattern matching techniques described above. We ran the experiment on the two machines with 2 (using 8 threads) and 4 Cores (using 64 threads) listed in Table \ref{tab:final-results}. There was no manual intervention during the complete run. The tool automatically generated the report after all the samples were processed.

\begin{table}[h!]
\centering
\footnotesize
{
\begin{tabular}{  c  c  c  c  c  c  c  } \hline\noalign{\smallskip}

\textbf{Experiment} & \textbf{Analysis} & \textbf{Detection} & \textbf{False}     & \textbf{Data Set Size}  & \textbf{Real-Time\footnotemark[1]} & \textbf{Platform\footnotemark[2]} \\
\textbf{Number}     & \textbf{Type}     & \textbf{Rate}      & \textbf{Positives} & \textbf{Benign/Malware} & \textbf{}                           & \textbf{} \\ \noalign{\smallskip}\hline\hline\noalign{\smallskip}

$1^{3,4}$ & Static   & 100\%         & 0\%          & 1137 / 250       & \ding{51}        & Win 32    \\ \noalign{\smallskip}\hline\noalign{\smallskip}
$2^{4,5}$ & Static   & 93.92\%       & 3.02\%       & 1137 / 250       & \ding{51}        & Win 32    \\ \noalign{\smallskip}\hline\noalign{\smallskip}
$2^{4,6}$ & Static   & 99.6\%        & 3.43\%       & 1137 / 250       & \ding{51}        & Win 32    \\ \noalign{\smallskip}\hline\noalign{\smallskip}
$2^{4,7}$ & Static   & 100\%         & 3.43\%       & 1137 / 250       & \ding{51}        & Win 32    \\ \noalign{\smallskip}\hline\noalign{\smallskip}

\end{tabular}
\\ [5pt]
\scriptsize
{
\begin{tabular}{  l  }
1 \;\;Real-time here means the detection is fully automatic and finishes in a reasonable amount of time. \\
2 \;\;All the samples (benigns and malwares) used were Windows 32 programs. \\ \\
\;\;\;\;\;Machines used in the experiments: \\ \\
3 \;\;With 2 Cores: \;\;\;\;\;\;\;Intel Core i5 CPU M 430 @ 2.27 GHz \\
\;\;\;\;\;RAM: \;\;\;\;\;\;\;\;\;\;\;\;\;\;\;\;\;\;\;\;4 GB \\
\;\;\;\;\;Operating System: Windows 8 Professional \\
4 \;\;With 4 Cores: \;\;\;\;\;\;\;Intel Core 2 Quad CPU Q6700 @ 2.67 GHz \\
\;\;\;\;\;RAM: \;\;\;\;\;\;\;\;\;\;\;\;\;\;\;\;\;\;\;\;4 GB \\
\;\;\;\;\;Operating System: Windows 7 Professional \\ \\
5 \;\;Results obtained using 10-fold cross validation \\
\;\;\;\;\;Training dataset: 25 samples Unknown malware samples: 225 \\
6 \;\;Training dataset: 100 samples, Unknown malware samples: 150 \\
7 \;\;Training dataset: 200 samples, Unknown malware samples: 50
\end{tabular}
}
}
\caption{\textbf{Results of the Experiments Carried out as Part of the Empirical Study}}
\label{tab:final-results}
\end{table}

\subsubsection{Experiment (2): To detect unknown malwares}

For this experiment we selected 25 malware samples out of the 250 malwares. These 25 malware samples were used to train the tool MARD to classify a program as benign or malware. These two steps were repeated 10 times and each time different set of 25 malware samples were selected for training (10-fold cross validation). After the program (sample) is translated to MAIL a CFG for each function in the program is build. Instead of using one large CFG as done in Experiment \textbf{(1)}, we divide a program into smaller CFGs. A program that contains some percenatge of the control flow of a training malware sample, can be classified as a malware. The CFGs of a program help the tool MARD to detect such unknown malwares as explained below.

\begin{table}[h!]
\centering
\footnotesize
{
\begin{tabular}{  c  c  c  } \hline\noalign{\smallskip}

\textbf{Threshold} & \textbf{Detection} & \textbf{False}      \\
\textbf{}          & \textbf{Rate}      & \textbf{Positives \footnotemark[1]}  \\ \noalign{\smallskip}\hline\hline\noalign{\smallskip}

10 \footnotemark[2]  & 100\%        & 3.07\%       \\ \noalign{\smallskip}\hline\noalign{\smallskip}
20                   & 99.2\%       & 3.07\%       \\ \noalign{\smallskip}\hline\noalign{\smallskip}
25 \footnotemark[3]  & 99.2\%       & 3.07\%       \\ \noalign{\smallskip}\hline\noalign{\smallskip}
30                   & 93.2\%       & 3.07\%       \\ \noalign{\smallskip}\hline\noalign{\smallskip}
40                   & 86.4\%       & 3.07\%       \\ \noalign{\smallskip}\hline\noalign{\smallskip}
50                   & 82.8\%       & 3.07\%       \\ \noalign{\smallskip}\hline\noalign{\smallskip}
60                   & 76\%         & 3.07\%       \\ \noalign{\smallskip}\hline\noalign{\smallskip}
70                   & 76\%         & 3.07\%       \\ \noalign{\smallskip}\hline\noalign{\smallskip}
80                   & 76\%         & 3.07\%       \\ \noalign{\smallskip}\hline\noalign{\smallskip}
90                   & 76\%         & 3.07\%       \\ \noalign{\smallskip}\hline\noalign{\smallskip}

\end{tabular}
\\ [5pt]
\scriptsize
{
\begin{tabular}{  l  }
1 \;\;The same set of training dataset is used for all the experiments, \\
\;\;\;\;\;therefore all of them have the same number of false positives. \\
2 \;\;We did not pick 10 as the threshold because we used only \\
\;\;\;\;\;one set of dataset and 100\% detection rate seems too perfect. \\
3 \;\;We picked 25 as the threshold because after this the \\
\;\;\;\;\;detection rate started falling considerably. \\ \\
Test dataset size: 1137 benigns and 250 malwares. \\
Training dataset size: 25 malwares. \\
Machine used in the experiment: \\
With 4 Cores: \;\;\;\;\;\;\;Intel Core 2 Quad CPU Q6700 @ 2.67 GHz \\
RAM: \;\;\;\;\;\;\;\;\;\;\;\;\;\;\;\;\;\;\;\;4 GB \\
Operating System: Windows 7 Professional \\ \\
\end{tabular}
}
}
\caption{\textbf{Results of the Experiment Carried out to Pick the Optimum Threshold Value to be Used in Experiment (2)}}
\label{tab:results-threshold}
\end{table}

Applying the graph and the pattern matching techniques described above the tool MARD matches CFGs of each 25 malware samples with the CFGs of a program to classify the program as either benign or malware. We base these classifications on a threashold value of 25\%. That is, if 25\% or more of the CFGs of a malware sample matches with the CFGs of a program then the program is classified as malware, else the program is classified as benign. The threshold value was computed by carrying out experiments with different range of threshold values as shown and explained in Table \ref{tab:results-threshold}. The results of experiment \textbf{(2)} are listed in Table \ref{tab:final-results}. The detection rate is 93.92\% because of the use of small number (25) of training dataset. The detection rate improved to 99.6\% and 100\% when we used a training dataset of 100 and 200 samples respectively.

\subsection{Limitations of MAIL}

A program translated to MAIL when executed may not produce the same output as the original program. The language MAIL is designed to perform static binary analysis and is not suitable for performing dynamic binary analysis.

The patterns developed if used with a behavioral signature of a binary program such as a CFG have the capability to produce useful classifications for malware analysis and detection, as shown by the results of the above experiments. But if the patterns are used alone, it may not produce the desired results.

The side effects of an assembly instruction is not directly translated to the MAIL statement. With the presence of various flag registers in the MAIL language it is possible for a malware analysis tool to include the side effect(s) of an assembly instruction by generating more statements and updating the affected flag registers.

The MAIL language is most useful in capturing the behavior (including structural and functional) of a binary program and can be used as part of different malware detection techniques such as described in this paper and in \cite{Code-Graph-MD, Mining-PE-MD, CFA-1, CFA-2}. These techniques require behavioral, structural or functional information about a program. In its current form the MAIL language cannot be used as part of other signature-based malware detection techniques, such as \cite{Bioinformatics-MD, Opcode-Graph-MD, Histogram-MD}. These techniques build the signatures using the opcodes of a binary program.

\section{Conclusion}\label{sec:conclusion}

We have developed the new language MAIL for malware anlaysis and have used it successfully in our tool MARD for malware analysis and detection. We carried out an experimental study and showed that we can achieve detection rates of: 100\% with 0\% false positives for known malwares and {\raise.17ex\hbox{$\scriptstyle\sim$}}(94 -- 100)\% with {\raise.17ex\hbox{$\scriptstyle\sim$}}(3 -- 3.5)\% false positives for unknown malwares, as shown in Table \ref{tab:final-results}. The two main contributions of the language MAIL are: \textbf{(1)} Providing platform independence and automation for malware analysis and detection tools, as is shown by its use in the tool MARD. \textbf{(2)} Optimizing the creation of a behavioral signature of a program, as is shown by creating a ACFG (Annotated Control Flow Graph), a CFG with patterns, of a binary program. We have shown how this ACFG is used for reliable malware analysis and detection in real-time. More recent examples of the use of MAIL can be found in \cite{MARD-COSE}.

Currently we are carrying out further research into optimizing the tool to increase its accuracy and efficiency for detecting unknown metamorphic malwares. We found that the behavioral signatures generated by the tool MARD using the MAIL language, and the graph and the pattern matching techniques, are helpful in detecting metamorphic malwares as shown in Table \ref{tab:final-results}. We are collecting more metamorphic malware samples to use in our research and carry out experiments to further improve malware classification and detection.

\bibliographystyle{plain}
\bibliography{mail}

\begin{thebibliography}{10}

\bibitem{Dragon-Book}
Alfred~V. Aho, Monica~S. Lam, Ravi Sethi, and Jeffrey~D. Ullman.
\newblock {\em Compilers: Principles, Techniques, and Tools (2nd Edition)}.
\newblock Addison-Wesley Longman Publishing Co., Inc., Boston, MA, USA, 2006.

\bibitem{MARD-COSE}
Shahid Alam, R~Nigel Horspool, Issa Traore, and Ibrahim Sogukpinar.
\newblock A framework for metamorphic malware analysis and real-time detection.
\newblock {\em Computers \& Security}, 48:212--233, February 2015.

\bibitem{ITU-Malware-Financial}
J.M. Bauer, M.J.G. Eeten, and Y.~Wu.
\newblock Itu study on the financial aspects of network security: Malware and
  spam.
\newblock {\em \copyright International Telecommunications Union
  (http://www.itu.int)}, 2008.

\bibitem{PARSEC}
Christian Bienia, Sanjeev Kumar, Jaswinder~Pal Singh, and Kai Li.
\newblock The parsec benchmark suite: Characterization and architectural
  implications.
\newblock In {\em Proceedings of the 17th international conference on Parallel
  architectures and compilation techniques}, PACT '08, pages 72--81, New York,
  NY, USA, 2008. ACM.

\bibitem{Virus-Undecidable-Problem-1}
F.~Cohen.
\newblock Computer viruses: Theory and experiments.
\newblock {\em Comput. Security.}, 6(1):22--35, Feburary 1987.

\bibitem{Cook-Isomorphism-NP-Complete}
Stephen~A. Cook.
\newblock The complexity of theorem-proving procedures.
\newblock In {\em Proceedings of the third annual ACM symposium on Theory of
  computing}, STOC '71, pages 151--158, New York, NY, USA, 1971. ACM.

\bibitem{Intel-Developer-Manual}
Intel Corporation.
\newblock {\em Intel \textregistered{} 64 and IA-32 Architectures Software
  Developer's Manual Volume 2 (2A, 2B \& 2C): Instruction Set Reference, A-Z},
  January 2013.

\bibitem{EBNF}
International Standard~Organization document~reference ISO/IEC.
\newblock {\em Information Technology - Syntactic Metalanguage - Extended
  Backus-Naur Form}, 14977 : 1996(E), 1996.

\bibitem{CFA-2}
Mojtaba Eskandari and Sattar Hashemi.
\newblock Ecfgm: Enriched control flow graph miner for unknown vicious infected
  code detection.
\newblock {\em Journal in Computer Virology}, 8(3):99--108, August 2012.

\bibitem{CFA-1}
Mojtaba Eskandari and Sattar Hashemi.
\newblock A graph mining approach for detecting unknown malwares.
\newblock {\em Journal of Visual Languages and Computing}, 23(3):154--162, June
  2012.

\bibitem{Mining-PE-MD}
Parvez Faruki, Vijay Laxmi, M.~S. Gaur, and P.~Vinod.
\newblock Mining control flow graph as api call-grams to detect portable
  executable malware.
\newblock In {\em Proceedings of the Fifth International Conference on Security
  of Information and Networks}, SIN '12, pages 130--137, New York, NY, USA,
  2012. ACM.

\bibitem{LLVM}
Chris Lattner and Vikram Adve.
\newblock Llvm: A compilation framework for lifelong program analysis \&
  transformation.
\newblock In {\em Proceedings of the international symposium on Code generation
  and optimization: feedback-directed and runtime optimization}, CGO '04,
  Washington, DC, USA, 2004. IEEE Computer Society.

\bibitem{Code-Graph-MD}
Jusuk Lee, Kyoochang Jeong, and Heejo Lee.
\newblock Detecting metamorphic malwares using code graphs.
\newblock In {\em Proceedings of the 2010 ACM Symposium on Applied Computing},
  SAC '10, pages 1970 -- 1977, New York, NY, USA, 2010. ACM.

\bibitem{Obfuscation-2}
Cullen Linn and Saumya Debray.
\newblock Obfuscation of executable code to improve resistance to static
  disassembly.
\newblock In {\em Proceedings of the 10th ACM conference on Computer and
  communications security}, CCS '03, pages 290--299, New York, NY, USA, 2003.
  ACM.

\bibitem{Virus-Undecidable-Problem-2}
David M.~Chess and Steve~R. White.
\newblock An undetectable computer virus.
\newblock {\em Virus Bulletin Conference}, September 2000.

\bibitem{Obfuscation-1}
Philip OKane, Sakir Sezer, and Kieran McLaughlin.
\newblock Obfuscation: The hidden malware.
\newblock {\em IEEE Security and Privacy}, 9(5):41--47, September 2011.

\bibitem{ARM-Architecture-Manual}
ARM~Holdings plc.
\newblock {\em ARM \textregistered{} Architecture Reference Manual ARMv7-A and
  ARMv7-R edition}, January 2012.

\bibitem{Histogram-MD}
B.B. Rad, M.~Masrom, and S.~Ibrahim.
\newblock Opcodes histogram for classifying metamorphic portable executables
  malware.
\newblock In {\em e-Learning and e-Technologies in Education (ICEEE), 2012
  International Conference on}, pages 209--213, sept. 2012.

\bibitem{Opcode-Graph-MD}
Neha Runwal, Richard~M. Low, and Mark Stamp.
\newblock Opcode graph similarity and metamorphic detection.
\newblock {\em J. Comput. Virol.}, 8(1-2):37--52, May 2012.

\bibitem{GCC}
GCC Team.
\newblock {\em GCC: The GNU Compiler Collection}.
\newblock http://gcc.gnu.org, 2013.

\bibitem{Bioinformatics-MD}
P.~Vinod, V.~Laxmi, M.S. Gaur, and G.~Chauhan.
\newblock Momentum: Metamorphic malware exploration techniques using msa
  signatures.
\newblock In {\em Innovations in Information Technology (IIT), 2012
  International Conference on}, pages 232--237, March 2012.

\end{thebibliography}

\appendix
\section{Appendix}\label{app:grammar-mail}

{
\small
\begin{verbatim}

TITLE:       MAIL (Malware Analysis Intermediate Language) Grammar in EBNF
AUTHOR:      Shahid Alam (salam@cs.uvic.ca)
DATED:       March 24, 2013
REVISION:    1.0 (March 24, 2013)

DESCRIPTION:

The grammar can be defined by a 3-tuple G = (T, N, P) where
T = set of terminals
N = set of non-terminals
P = set of production rules

This document describes the grammar for MAIL. The grammar uses the EBNF syntax, where
'|' means a choice, ? means optional, * means zero or more times and + means one or more
times. Line Comments start with "--". Terminator symbol is ";". Terminals are enclosed
in single quotes.

----------------------------------------------------------------------------------------
--                                    PRODUCTION RULES                                --
----------------------------------------------------------------------------------------

statements          ::= ( statement* ) ;
statement           ::= assignment_s+
                        | control_s+
                        | condition_s+
                        | function_s+
                        | jump_s+
                        | lib_call_s+
                        | 'halt'
                        | 'lock' ;


assignment_s        ::= register_s
                        | address_s ;

register_s          ::= register '=' (math_operator)? expr
                        | register '=' (expr)? math_operator expr
                        | register '=' lib_call_s ;

address_s           ::= address '=' (math_operator)? expr
                        | address '=' (expr)? math_operator expr
                        | address '=' lib_call_s ;

control_s           ::= ( 'if' condition_s (jump_s | assignment_s) )
                        ( 'else' (jump_s | assignment_s) )? ;

jump_s              ::= 'jmp' address ;

lib_call_s          ::= letter+ '(' address (, args)* ')' ;

function_s          ::= 'start_function_' digit+ statement 'end_function_' digit+ ;

condition_s         ::= (expr rel_operator expr)+ ;

----------------------------------------------------------------------------------------
--                                      HELPER RULES                                  --
----------------------------------------------------------------------------------------

expr                ::= register
                        | address
                        | digit+ ;

register            ::= 'eflags'
                        | 'gr_' digit+
                        | 'fr_' digit+
                        | 'sp'
                        | register_name (':' register_name)? ;

register_name       ::= letter+ ['0' - '9']? ;

address             ::= '[' digit+ ']'
                        | reg_address
                        | 'UNKNOWN' ;

reg_address         ::= '[' register ( arith_operator (register | digit+) )* ']'
                        | '[' sp '=' sp ('+' | '-') digit+ ']'
                        | '[' register (':' register)? ']' ;

letter              ::= ['a' - 'z'] ['A' - 'Z'] ;

digit               ::= '0x' ['0' - '9'] | ['A' - 'F'] ;

math_operator       ::= arith_operator | log_operator ;

arith_operator      ::= '+' | '-' | '*' | '/' | '%' | '.' ;

log_operator        ::= 'and' | 'or' | 'xor' | ! | '<<' | '>>' ;

args                ::= address (',' address)* ;

rel_operator        ::= '<' | '>' | '<=' | '>=' | '==' | '!=' ;

comment             ::= '--' blank | tab | character | comment* newline ;

character           ::= '!' | '"' | '#' | '$' | '%' | '&' | ''' | '(' | ')'
                        | '[' | '\' | ']' | '^' | '_' | '`' | '{' | '|' | '}'
                        | '*' | '+' | '-' | '/' | ',' | '.' | '~'
                        | ':' | ';' | '<' | '=' | '>' | '?' | '@'
                        | ['0' - '9']	| letter ;

----------------------------------------------------------------------------------------
--                                          TOKENS                                    --
----------------------------------------------------------------------------------------

WS                  ::= blank | tab | newline ;
COMMENT             ::= '--' blank | tab | character | comment* newline ;
NUM                 ::= digit+ ;
COMMA               ::= ',' ;
COLON               ::= ':' ;
SCOLON              ::= ';' ;
LOP                 ::= 'and' | 'or' | 'xor' | ! | '<<' | '>>' ;
AOP                 ::= '+' | '-' | '*' | '/' | '%' | '.' ;
ROP                 ::= '<' | '>' | '<=' | '>=' | '==' | '!=' ;
SFUN                :: 'start_function_' digit+ ;
EFUN                :: 'end_function_'  digit+ ;
EQUAL               :: '=' ;
MUL                 ::= '*' ;
DIV                 ::= '/' ;
PLUS                ::= '+' ;
MINUS               ::= '-' ;
LBRKT1              ::= '(' ;
RBRKT1              ::= ')' ;
LBRKT2              ::= '[' ;
RBRKT2              ::= ']' ;
IF                  ::= 'if' ;
ELSE                ::= 'else' ;
UNKNOWN             ::= 'UNKNOWN' ;

\end{verbatim}
}

\section{Appendix}\label{app:ignored-instructions}

List of x86 and ARM instructions, in alphabetical order, that are not translated to MAIL statements: \\

\noindent
\textbf{\underline{x86}}

{
\small
\begin{verbatim}
3DNOW
AAA, AAD, AAM, AAS, AESDEC, AESDECLAST, AESENC
AESENCLAST, AESIMC, AESKEYGENASSIST, ARPL
BOUND
DAA, DAS
EMMS, ENTER
GETSEC
CLFLUSH, CLTS, CMC, CPUID, CRC32
FCLEX, FDECSTP, FEDISI, FEEMS, FENI, FFREE, FINCSTP, FINIT
FLDCW, FLDENV, FNCLEX, FNINIT, FNSAVE, FNSTCW, FNSTENV
FNSTSW, FRSTOR, FSAVE, FSETPM, FSTCW, FSTENV, FSTSW
FXRSTOR, FXRSTOR64, FXSAVE, FXSAVE64, FXTRACT
INT 3, INVD, INVEPT, INVLPG, INVLPGA, INVPCID, INVVPID
LEAVE, LFENCE, LZCNT
MFENCE, MONITOR, MPSADBW, MWAIT
PAUSE, PREFETCH, PSAD, PSHUF, PSIGN
RCL, RCR, RDRAND, RDTSC, RDTSCP, ROL, ROR, RSM
SFENCE, SHUFPD, SHUFPS, SKINIT, SMSW
SYSCALL, SYSENTER, SYSEXIT, SYSRET
VAESDEC, VAESDECLAST, VAESENC, VAESENCLAST, VAESIMC
VAESKEYGENASSIST, VERR, VERW, VMCALL, VMCLEAR, VMFUNC
VMLAUNCH, VMLOAD, VMMCALL, VMPSADBW, VMREAD, VMRESUME
VMRUN, VMSAVE, VMWRITE, VMXOFF, VSHUFPD, VSHUFPS
VZEROALL, VZEROUPPER
WAIT, WBINVD
XRSTOR, XSAVE, XSAVEOPT
\end{verbatim}
}

\noindent
\textbf{\underline{ARM}}

{
\small
\begin{verbatim}
BKPT
CDP, CDP2, CLREX, CLZ, CPS, CPSID, CPSIE, CRC32, CRC32C
DBG, DCPS1, DCPS2, DCPS3, DMB, DSB
HVC
ISB
LDC, LDC2
MCR, MCR2, MCRR, MCRR2, MRC, MRC2, MRRC, MRRC2
PLD, PLDW, PLI, PRFM
SETEND, SEV, SHA1C, SHA1H, SHA1M, SHA1P, SHA1SU0, SHA1SU1
SHA256H, SHA256H2, SHA256SU0, SHA256SU1, SMC
SSAT, SSAT16, STC, STC2, SVC
USAT, USAT16
VCVT, VCVTA, VCVTB, VCVTM, VCVTN, VCVTP, VCVTR, VCVTT, VEXT
VLD1, VLD2, VLD3, VLD4, VLD1, VLD2, VLD3, VLD4
VQRDMULH, VQRSHL, VQRSHRN, VQRSHRN, VQRSHRUN, VQRSHRUN,
VQSHL, VQSHLU, VQSHL, VQSHRN, VQSHRN, VQSHRUN, VQSHRUN
VRINTA, VRINTM, VRINTN, VRINTP, VRINTR, VRINTX, VRINTZ
VRSHL, VRSHR, VRSHRN, VRSRA, VRSUBHN
VST1, VST2, VST3, VST4, VTBL, VTBX, VTRN, VUZP, VZIP
WFE, WFI, YIELD
\end{verbatim}
}

\end{document}